\documentclass{jfm}
\usepackage{graphicx}
\usepackage{amsbsy}
\usepackage{amsmath}
\usepackage{subfig}
\usepackage{natbib}
\usepackage{color}
\newcommand{\mycite}[1]{\cite[][]{#1}}
\newcommand{\citeasnoun}[1]{\cite{#1}}
\newcommand{\citenameasnoun}[1]{\citeasnoun{#1}}
\newcommand{\mysection}{\S\,}
\newcommand{\opd}{\mathrm d}
\newcommand{\opi}{\mathrm i}
\newcommand{\ope}{\mathrm e}
\renewcommand{\vec}[1]{\boldsymbol #1}
\newcommand{\matrx}[1]{\mathsfbi{#1}}
\newcommand{\taun}{\myupper { \mathsfbi{\tau}}{n}}
\newcommand{\taunpa}{\myupper { \mathsfbi{\tau}}{n+1}}
\renewcommand{\Im}{\operatorname{Im}}
\newcommand{\figref}[1]{figure~\ref{fig:#1}}

\newcommand{\eqnumref}[1]{(\ref{eq:#1})}
\renewcommand{\eqref}[1]{(\ref{eq:#1})}
\newcommand{\eqreftwo}[2]{equations~\eqnumref{#1} and~\eqnumref{#2}}
\newcommand{\myupper}[2]{#1^{(#2)}}
\newcommand{\un}{\myupper{\vec{u}}{n}}
\newcommand{\unpa}{\myupper{\vec{u}}{n+1}}
\newcommand{\urn}{\myupper{u_r}{n}}
\newcommand{\uzn}{\myupper{u_z}{n}}
\newcommand{\pn}{\myupper{p}{n}}
\newcommand{\Rn}{\myupper{R}{n}}
\newcommand{\Rnma}{\myupper{R}{n-1}}

\newcommand{\mun}{\myupper{\mu}{n}}
\newcommand{\munpa}{\myupper{\mu}{n+1}}
\newcommand{\zetan}{\myupper{\zeta}{n}}
\newcommand{\gamman}{\myupper{\gamma}{n}}
\newcommand{\phin}{\myupper{\phi}{n}}
\newcommand{\deltaRn}{\myupper{\delta R}{n}}
\newcommand{\deltaurn}{\myupper{\delta u_r}{n}}
\newcommand{\deltauzn}{\myupper {\delta u_z}{n}}
\newcommand{\deltapn}{\myupper {\delta p}{n}}
\newcommand{\cna}{\myupper {c_1}{n}}
\newcommand{\cnb}{\myupper {c_2}{n}}
\newcommand{\cnc}{\myupper {c_3}{n}}
\newcommand{\cnd}{\myupper {c_4}{n}}

\newcommand{\matBn}{\myupper {\matrx{B}}{n}}

\newcommand{\nn}{\myupper{\vec{n}}{n}}
\newcommand{\tn}{\myupper{\vec{t}}{n}}
\newcommand{\kappan}{\myupper{\kappa}{n}}
\newcommand{\Kzn}{\myupper{K_0}{n}}
\newcommand{\Izn}{\myupper{I_0}{n}}
\newcommand{\Kan}{\myupper{K_1}{n}}
\newcommand{\Ian}{\myupper{I_1}{n}}
\newcommand{\MN}{\myupper {\matrx{M}}{N}} 
\newcommand{\MaN}{\myupper {\matrx{M}_1}{N}}
\newcommand{\MbN}{\myupper {\matrx{M}_2}{N}}
\newcommand{\sigmamax}{\sigma_{\max}}
\newcommand{\phibarn}{\myupper{\bar{\phi}}{n}}

\newcommand{\myS}{\mathcal{S}}
\newcommand{\myE}{\mathcal{E}}
\newcommand{\deltaRnj}{\myupper{\delta R_j}{n}}
\newcommand{\bolddeltaRj}{{\boldsymbol {\delta R}_j}}
\newcommand{\bolddeltaRmax}{\boldsymbol {\delta R}_{\rm max} }
\newcommand{\bolddeltaRmaxeff}{\boldsymbol {\delta R}_{\rm max}^{\rm eff}}

\newcommand{\muin}{\mu_{\rm in}}
\newcommand{\mumid}{\mu_{\rm mid}}
\newcommand{\muout}{\mu_{\rm out}}
\newcommand{\Rin}{R_{\rm in}}
\newcommand{\Rout}{R_{\rm out}}
\newcommand{\gammacont}{\eta_{\rm cont}}
\newcommand{\vecnabla} {\boldsymbol{\nabla}}
\newcommand{\vecx}{\vec{x}}
\newcommand{\kmax}{k_{\rm max}}

\newcommand{\sigmaeff}{\sigma^{\rm eff}}
\newcommand{\sigmaeffmax}{\sigmaeff_{\rm max}}

\newcommand{\matAnnp}{\myupper{\matrx{A}}{n,n'}}
\newcommand{\matAnpn}{\myupper{\matrx{A}}{n',n}}
\newcommand{\munp}{\myupper{\mu}{n'}}
\newcommand{\matAnn}{\myupper{\matrx{A}}{n,n}}
\newcommand{\matAnnpa}{\myupper{\matrx{A}}{n,n+1}}

\newcommand{\rhon}{\myupper{\rho}{n}}
\newcommand{\rhonpr}{\myupper{\rho}{n'}}
\newcommand{\kn}{\myupper{k}{n}}

\newcommand{\munpr}{\myupper{\mu}{n'}}

\newcommand{\Ka}{K_1}
\newcommand{\Kz}{K_0}
\newcommand{\Ia}{I_1}
\newcommand{\Iz}{I_0}
\newcommand{\knpr}{\myupper{k}{n'}}

\newcommand{\alphanpr}{\myupper{\alpha}{n'}}
\newcommand{\lambdan}{\myupper{\lambda}{n}}
\newcommand{\lambdanpr}{\myupper{\lambda}{n'}}
\newcommand{\myfontsize}{\fontsize{9}{11}\selectfont}
\newcommand{\baruzn}{\myupper{\bar{u}_z}{n}}
\newcommand{\baruznp}{\myupper{\bar{u}_z}{n'}}
\newcommand{\barvz}{\bar{v}_z}
\newcommand{\Kznpr}{\myupper{K_0}{n'}}
\newcommand{\Iznpr}{\myupper{I_0}{n'}}
\newcommand{\Kanpr}{\myupper{K_1}{n'}}
\newcommand{\Ianpr}{\myupper{I_1}{n'}}

\providecommand{\noopsort}[1]{}

\begin{document}

\title[Stability analysis of concentric shells]{Linear stability analysis of capillary instabilities for concentric cylindrical shells}

\author[X. Liang, D. S. Deng, J.-C. Nave, and S. G. Johnson]%
{X.\ns L\ls I\ls A\ls N\ls G$^1$,\ns 
D.\ns S.\ns D\ls E\ls N\ls G$^2$,\break
J.\ls-\ls C.\ns N\ls A\ls V\ls E$^3$\ns
\and S\ls T\ls E\ls V\ls E\ls N\ns G.\ns J\ls O\ls H\ls N\ls S\ls O\ls N$^1$}

\affiliation{$^1$Department of Mathematics, Massachusetts Institute of Technology, Cambridge, MA 02139, USA\\[\affilskip]
$^2$Department of Chemical Engineering, Massachusetts Institute of Technology, Cambridge, MA 02139, USA\\[\affilskip]
$^3$Department of Mathematics and Statistics, McGill University, Montreal, Quebec, Canada}

\maketitle
\date{}
\begin{abstract}
Motivated by complex multi-fluid geometries currently being explored in fibre-device manufacturing, we study capillary instabilities in concentric cylindrical flows of $N$ fluids with arbitrary viscosities, thicknesses, densities, and surface tensions in both the Stokes regime and for the full Navier--Stokes problem. Generalizing previous work by Tomotika ($N=2$), Stone \& Brenner ($N=3$, equal viscosities) and others, we present a full linear stability analysis of the growth modes and rates, reducing the system to a linear generalized eigenproblem in the Stokes case. Furthermore, we demonstrate by Plateau-style geometrical arguments that only axisymmetric instabilities need be considered. We show that the $N=3$ case is already sufficient to obtain several interesting phenomena: limiting cases of thin shells or low shell viscosity that reduce to $N=2$ problems, and a system with competing breakup processes at very different length scales. The latter is demonstrated with full 3-dimensional Stokes-flow simulations. Many $N > 3$ cases remain to be explored, and as a first step we discuss two illustrative $N \to \infty$ cases, an alternating-layer structure and a geometry with a continuously varying viscosity.
\end{abstract}


\section{Introduction}
In this paper, we generalize previous linear stability analyses~\mycite{Rayleigh1879, Rayleigh1892, Tomotika1935, Stone1996, Chauhan2000} of Plateau--Rayleigh (capillary) instabilities in fluid cylinders to handle any number ($N$) of concentric cylindrical fluid shells with
arbitrary thicknesses, viscosities, densities, and surface tensions.  This
analysis is motivated by the fact that experimental work is currently
studying increasingly complicated fluid systems for device-fabrication
applications, such as drawing of microstructured optical fibres with
concentric shells of different glasses/polymers~\mycite{ Hart2002, Kuriki2004, Pone2006, Abouraddy2007, Sorin2007, Deng2008} or generating double emulsions~\mycite{Utada2005, Shah2008}. Although real experimental geometries may not be exactly concentric, we show that surface tension alone, in the absence of other forces, will tend to eliminate small deviations from concentricity.  We show that
our solution reduces to known results in several limiting cases, and
we also validate it with full 3-dimensional Stokes-flow simulations.  In addition, we show
results for a number of situations that have not been previously
studied. For the limiting case of a thin shell, we show a connection
to the classic single-cylinder and flat-plane results, consistent with a similar result for air-clad two-fluid jets~\mycite{Chauhan2000}.  In a
three-fluid system, we exhibit an interesting case in which two growth
modes at different wavelengths have the same effective growth rate,
leading to competing breakup processes that we demonstrate with full
3-dimensional Stokes-flow simulations. We also consider some many-layer cases, including a
limiting situation of a continuously varying viscosity.  Using a
simple geometrical argument, we generalize previous results~\mycite{Plateau1873, Rayleigh1879, Chandrasekhar1961} to
show that only axial (not azimuthal) instabilities need be considered
for cylindrical shells. Numerically, we show that the stability
analysis in the Stokes regime can be reduced to a generalized eigenproblem whose solutions
are the growth modes, which is easily tractable even for large numbers
of layers. Like several previous authors~\mycite{Tomotika1935, Stone1996, Gunawan2002, Gunawan2004}, we begin by considering the Stokes
(low-Reynolds) regime, which is consistent with the high viscosities
found in drawn-fibre devices~\mycite{Abouraddy2007,Deng2008}. In Appendix~\ref{SecFullNS}, we generalize the analysis to the full incompressible Navier-Stokes problem, which turns out to be a relatively minor modification once the Stokes problem is understood, although it has the complication of yielding an unavoidably nonlinear eigenproblem for the growth modes. Semi-analytical methods are a crucial complement to large-scale Navier-Stokes simulations (or experiments) in studying capillary instabilities, since the former allow rapid exploration of wide parameter regimes (e.g. for materials design) as well as rigorous asymptotic results, while the latter can capture the culmination of the breakup process as it grows beyond the linear regime.

Capillary instability of liquid jets has been widely
studied~\mycite{Lin2003, Eggers2008} since \citenameasnoun{Plateau1873} first
showed that whenever a cylindrical jet's length exceeds its
circumference, it is always unstable due to capillary forces (surface
tension). Plateau used simple geometrical arguments based on comparing
surface energies before and after small perturbations. Lord Rayleigh
introduced the powerful tool of linear stability analysis and
reconsidered inviscid water jets~\mycite{Rayleigh1879} and viscous
liquid jets~\mycite{Rayleigh1892}. In linear stability analysis, one
expands the radius $R$ as a function of axial coordinate $z$ in the
form $R(z) = R_0 + {\delta R} \ope^{\opi kz - \opi \omega t}$, where $\delta R \ll
R_0$, $2\pi/k$ is a wavelength of the instability, and $\sigma = \Im
\omega$ is an exponential growth rate. Given a geometry, one solves
for the dispersion relation(s) $\omega(k)$ and considers the most
unstable growth mode with the growth rate $\sigmamax$ to determine the time scale of the
breakup process.  The wavelength $2\pi/k$ corresponding to $\sigmamax$
has been experimentally verified to match the disintegration size
of liquid jets~\mycite{Eggers2008}. By considering the effect of the
surrounding fluid, \citenameasnoun{Tomotika1935} generalized this
analysis to a cylindrical viscous liquid surrounded by another viscous
fluid, obtaining a $4 \times 4$ determinant equation for the
dispersion relation. (Rayleigh's results are obtained in the limit of
vanishing outer viscosity.)  A few more limiting solutions have been
obtained ~\mycite{Meister1967,Kinoshita1994} by generalizing
Tomotika's approach to non-Stokes regimes.  Beyond the regime of a
single cylindrical jet, \citenameasnoun{Stone1996} analysed the
three-fluid ($N=3$) Stokes cylinder problem, but only for equal
viscosities. \citeasnoun{Chauhan2000} analysed the $N=3$ case where the inner two fluids have arbitrary viscosities and the outermost fluid is inviscid gas, taking into account the full Navier--Stokes equations. \citenameasnoun{Gunawan2002,Gunawan2004} considered an array of
identical viscous cylinders (in a single row or in a triangular
configuration).

Much more complicated, multi-fluid geometries are now being considered
in experimental fabrication of various devices by a fibre-drawing
process~\mycite{ Hart2002, Kuriki2004, Pone2006, Abouraddy2007, Sorin2007,Deng2008}.  In fibre drawing, a scale model (preform) of
the desired device is heated to a viscous state and then pulled
(drawn) to yield a long fibre with (ordinarily) identical cross-section but much
smaller diameter.  For example, concentric layers of different
polymers and glasses can be drawn into a long fibre with
submicron-scale layers that act as optical devices for wavelengths on
the same scale as the layer thicknesses~\mycite{Abouraddy2007}.  Other
devices, such as photodetectors~\mycite{Sorin2007}, semiconductor
filaments~\mycite{Deng2008,Deng2010b}, and piezoelectric pressure sensors~\mycite{Egusa2010} have similarly been incorporated into microstructured fibre devices.  That
work motivates greater theoretical investigation of multi-fluid
geometries, and in particular the stability (or instability time scale)
of different geometries is critical in order to predict whether they
can be fabricated successfully.  For example, an interesting azimuthal
breakup process was observed experimentally~\mycite{Deng2008} and has
yet to be explained~\mycite{Deng2010}; however, we show in this paper
that azimuthal instability does not arise in purely cylindrical
geometries and must stem from the rapid taper of the fibre radius from centimetres to millimetres (the drawn-down ``neck''), or some other physical influence. [For example, there may be elastic effects (since fibres are drawn under tension), thermal gradients at longer length scales in the fibre (although breakup occurs at small scales where temperatures are nearly uniform), and long-range (e.g., van der Waals) interactions in submicron-scale films.] Aside from fibre drawing, recent authors have investigated multi-fluid microcapillary devices, in which the instabilities are exploited to generate double emulsions, i.e. droplets within droplets~\mycite{Utada2005}. Because the available theory was limited to equal viscosities, the experimental researchers chose only fluids in that regime, whereas our paper opens the possibilities of predictions for unequal viscosity and more than three fluids.

\begin{figure}
  \centering
  \subfloat[Azimuthal cross-section]{\label{fig:FigAzimuthal}\includegraphics[width=0.36\textwidth]{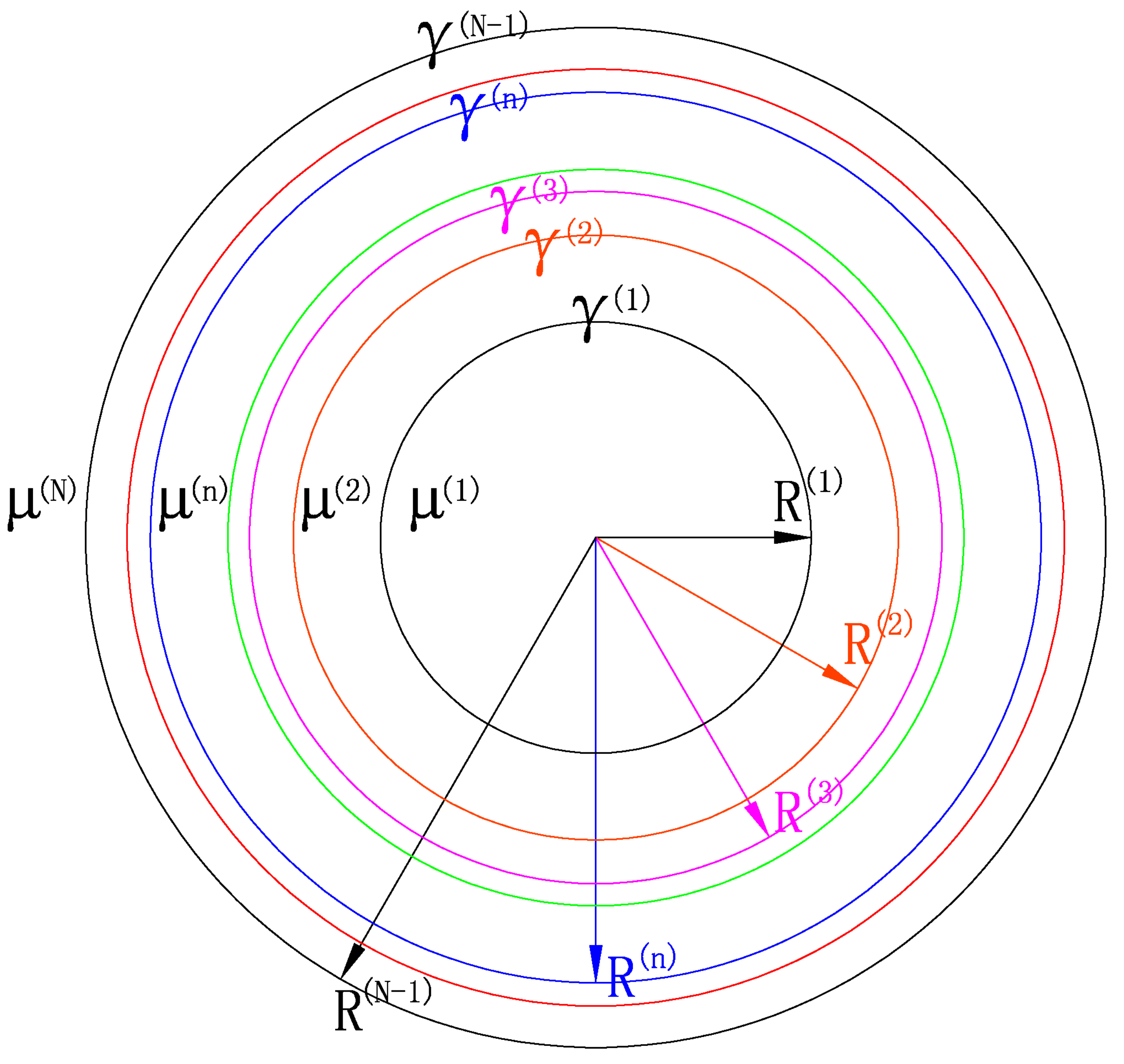}}                
  \subfloat[Steady state]{\label{fig:FigSteady}\includegraphics[width=0.3\textwidth,height=50mm]{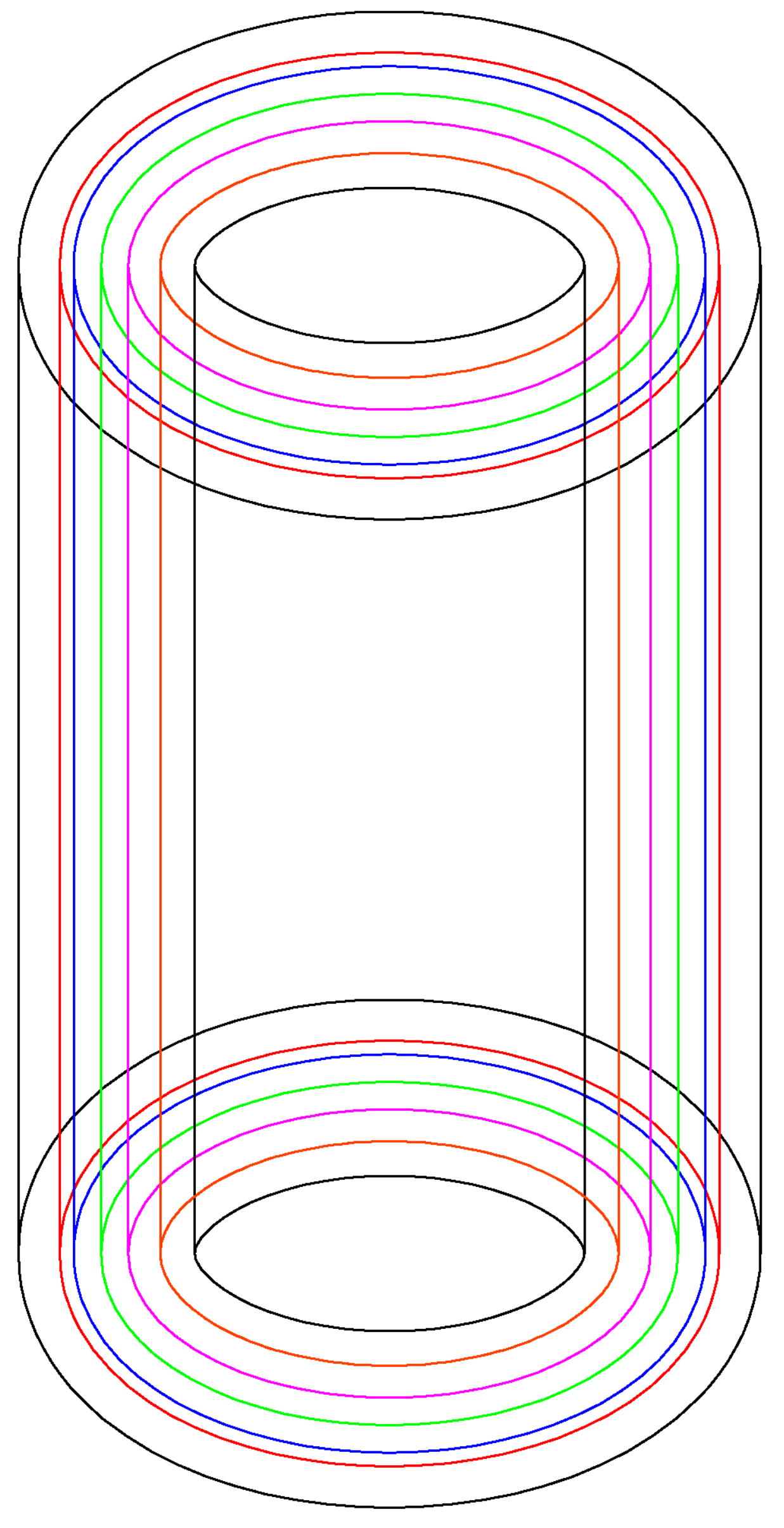}}
  \subfloat[Perturbed state]{\label{fig:FigPerturbed}\includegraphics[width=0.3\textwidth,height=50mm]{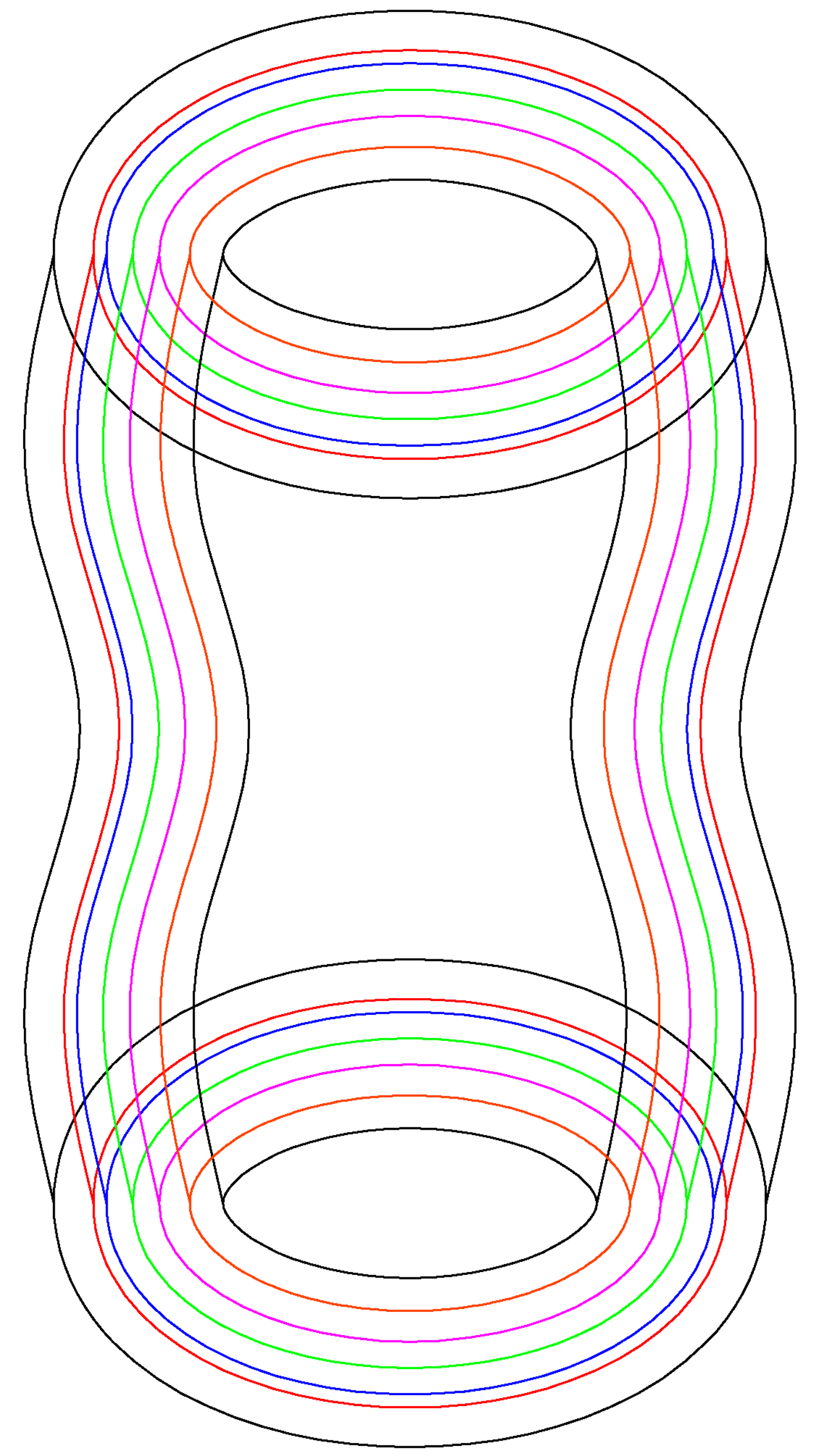}}
  \caption{Schematic of the concentric-cylinder geometry considered in this paper. (a) Cross-section of $N$ layers and corresponding radii $\Rn$, viscosities $\myupper{\mu}{n}$, and surface tensions $\myupper{\gamma}{n}$. Starting with the perfect cylindrical geometry (b), we then introduce small sinusoidal perturbations (c) and analyse their growth with linear stability analysis.}
  \label{fig:FigGeometry}
\end{figure}

We now formulate the mathematical problem that we solve, as depicted
schematically in \figref{FigGeometry}.  The total number of viscous
fluids is $N$ and the viscosity of the $n$-th ($n=1, 2, \cdots, N$) fluid
is $\mun$.  The surface-tension coefficient between the $n$-th and
$(n+1)-\mathrm{th}$ fluid is denoted by $\gamman$. For the unperturbed steady
state (\figref{FigSteady}), we assume that the $n$-th fluid is in a
cylindrical shell geometry with outer radius $ \Rn $ and inner radius
$\myupper{R}{n-1} < \Rn $. The first ($n=1$) fluid is the innermost
core and the $N$-th fluid is the outermost one (extending to
infinity), so we set $\myupper{R}{0} = 0$ and $\myupper{R}{N} = +
\infty$. To begin with, this system is analysed in the Stokes regime (low Reynolds
number) and we also neglect gravity [in the large Froude number
limit, valid for fibre-drawing~\mycite{Deng2011}], so the fluid densities are irrelevant. In Appendix~\ref{SecFullNS}, we extend this analysis to the Navier--Stokes regime, including an inertia term for each layer (with density $\rhon$).  As noted above, linear stability analysis consists of perturbing each 
interface $\Rn$ by a small sinusoidal amount $\deltaRn \ope^{\opi kz -\opi \omega t}$, to lowest order in $\deltaRn$. Stokes' equations are
then solved in each layer in terms of Bessel functions, and matching
boundary conditions yields a set of equations relating $\omega$ and
$k$.  Although these equations can be cast in the form of a polynomial
root-finding problem, similar to \citenameasnoun{Tomotika1935}, such a
formulation turns out to be ill-conditioned for large $N$, and instead
we formulate it in the Stokes regime as a generalized eigenproblem of the form
$\matrx{M}_2(k) \boldsymbol{\xi} = \opi \omega \matrx{M}_1(k)
\boldsymbol{\xi}$, which is easily solved for the dispersion relations
$\omega(k)$ (with the corresponding eigenvectors $\boldsymbol{\xi}$
yielding the relative amplitudes of each layer). In the Navier--Stokes regime, this becomes a nonlinear eigenproblem.
 
\section{Azimuthal stability}\label{SecStableOrNot}
For any coupled-fluids system of the type described in \figref{FigGeometry}, a natural question to ask is whether that system is stable subject to a small perturbation. If an interface with area $\myS$ has surface energy $\gamma \myS$, then a simple way to check stability is to compare surface energies (areas) for an initial configuration and a slightly deformed configuration with the same volume. In this way, it was shown that any azimuthal deformation is stable for a single cylindrical jet~\mycite{Rayleigh1879,Chandrasekhar1961}. Here, we employ a similar approach to demonstrate that the same property also holds for multiple concentric cylindrical shells. Note that this analysis only indicates whether a system is stable; in order to determine the time scale of an instability, we must use linear stability analysis as described in subsequent sections.

For the unperturbed system, we define the level-set function $\phibarn = r - \Rn$. $\phibarn = 0$ defines the interface between the $n$-th and $(n+1)$-th fluids. Similarly, we define the level-set function for the perturbed interface (\figref{FigPerturbed}) between the $n$-th and $(n+1)$-th fluids by
\begin{equation}
  \label{eq:eqLevelSet2a}
  \phin(r,z,\phi) = r - \zetan(z,\phi).
\end{equation}
 Following the method of normal modes~\mycite{Drazin2004}, in the limit of small perturbations, a disturbed interface $\zetan$ can be chosen in the form
\begin{equation}\label{eq:eqLevelSet2b}
    \zetan (z,\phi)  = \Rn  + \deltaRn \ope^{\opi(k z + m\phi)} + O[(\deltaRn)^2], 
\end{equation}
Assuming incompressible fluids in each layer so that volume is conserved (and assuming that the cylinder is much longer than its diameter so that any inflow/outflow at the end facets is negligible), we obtain a relation between $\zetan$ and $\Rn$
\begin{equation}
  \label{eq:eqRandRbar}
  \zetan(z,\phi) = \Rn + \deltaRn \ope^{\opi(k z + m\phi)} -\frac{(\deltaRn)^2}{4\Rn} + O[(\deltaRn)^3].
\end{equation}
Let $\myS (\phin)$ denote the surface area of $\phin(r,z,\phi) = 0$ in one wavelength $2\pi/k$. From \eqreftwo{eqLevelSet2a}{eqLevelSet2b}, $\myS (\phin)$  can be expressed in cylindrical coordinates as:
\begin{equation}
  \label{eq:eqSurfaceArea}
  \myS(\phin) = \int_0^{\frac{2\pi}{k}} \int_0^{2 \pi} \zetan(z,\phi) \sqrt{1+ (\frac{\partial \zetan}{\partial z})^2 + (\frac{1}{\zetan} \frac{\partial \zetan}{\partial \phi})^2} \opd \phi \opd z.
\end{equation}
Now we can compare the total interfacial energy between the unperturbed system  $\bar{\myE} = \sum_{n=1}^{N-1} \gamman \myS(\phibarn)$ and the perturbed system $\myE = \sum_{n=1}^{N-1} \gamman \myS(\phin)$: 
\begin{equation}
  \label{eq:eqEnergyCompare}
  \begin{split}
 \myE - \bar{\myE} & =  \sum_{n=1}^{N-1} \gamman \left[ \myS(\phin) -  \myS(\phibarn) \right]  \\
& =  \sum_{n=1}^{N-1} \gamman (\deltaRn)^2\pi^2 \frac{(k\Rn)^2 + m^2 -1}{k\Rn} \ + O[(\deltaRn)^3].
  \end{split}
\end{equation}
From the surface-energy point of view, small perturbations will grow only if $ \myE - \bar{\myE} < 0$. Therefore, from \eqref{eqEnergyCompare}, we can conclude that all the non-axisymmetric perturbations $(m \neq 0)$ will be stable. There is one special case that needs additional consideration: if $k=0$ and $m=1$ in \eqref{eqEnergyCompare}, the first term is zero, so one must consider the next-order term in order to show that this case is indeed stable (i.e. elliptical perturbations decay). Even more straightforwardly, however, $k=0$ corresponds to a two-dimensional problem, in which case it is well known that the minimal surface enclosing a given area is a circle. 

\section{Linear stability analysis}
\label{SecMainAnalysis}
In the previous section, we showed that only axisymmetric perturbations can lead to instability of concentric cylinders. Now we will use linear stability analysis to find out how fast the axisymmetric perturbations grow and estimate the break up time scale for a coupled $N$-layer system.

Here, we consider fluids in the low-Reynolds-number regime [valid for fibre-drawing \mycite{ Abouraddy2007, Deng2008}] and thus the governing equations of motion for each fluid are the Stokes equations~\mycite{Ockendon1995}. The full Navier--Stokes equations are considered in Appendix~\ref{SecFullNS}. For the axisymmetric flow, the velocity profile  of the $n$-th fluid is $\un = [\urn(r,z,t), \uzn(r,z,t)]$, where $\urn$  is the radial component of the velocity and $\uzn$ is the axial component of the velocity. The dynamic pressure in the $n$-th fluid is denoted by $\pn$. The Stokes equations~\mycite{Batchelor1973} are
\begin{subequations}
  \label{eq:eqStokes}
\begin{align}
\mun \left( \frac{\partial^2 \urn}{  \partial r^2} + \frac{1}{r} \frac{\partial \urn}{\partial r} - \frac{\urn}{r^2} + \frac{\partial^2 \urn}{\partial z^2} \right) = \frac{ \partial \pn}{ \partial r} \label{eq:eqStokesr} \\
\mun \left( \frac{\partial^2 \uzn}{  \partial r^2} + \frac{1}{r} \frac{\partial \uzn}{\partial r} + \frac{\partial^2 \uzn}{\partial z^2} \right) = \frac{ \partial \pn}{ \partial z} \label{eq:eqStokesz} 
\end{align}
\end{subequations}
and the continuity equation (for incompressible fluids) is
\begin{equation}\label{eq:eqContinuity}
 \frac{\partial \urn}{\partial r} + \frac{\urn}{r} + \frac{\partial \uzn}{\partial z} = 0.
\end{equation}

\subsection{Steady state}\label{sec:SteadyState}
 Because of the no-slip boundary conditions of viscous fluids, without loss of generality, we can take the equilibrium state of the outermost fluid to be 
\begin{equation}\label{eq:eqb4N}
\myupper{\bar{u}_r}{N} = 0 \qquad \myupper{\bar{u}_z}{N} = 0 \qquad \myupper{\bar{p}}{N} = 0
\end{equation}
for $r > \myupper{R}{N-1}$, and of the $n$-th $(n<N)$ fluid to be
\begin{equation}\label{eq:eqb4}
\myupper{\bar{u}_r}{n} = 0 \qquad \myupper{\bar{u}_z}{n} = 0 \qquad \myupper{\bar{p}}{n}  = \sum_{j=n}^{N-1} \frac{\myupper{\gamma}{j}}{\myupper{R}{j}}
\end{equation}
for $\Rnma<r <\Rn$. (In Appendix~\ref{SecFullNS}, we generalize this to nonzero initial relative velocities for the case of inviscid fluids, where no-slip boundary conditions are not applied.)

\subsection{Perturbed state}\label{sec:PerturbedState}
The perturbed interface, corresponding to the level set $\phin = r - \zetan = 0$, with an axisymmetric perturbation, takes the form
\begin{equation}\label{eq:eqb5}
    \zetan (z,t)  =  \Rn  + \deltaRn \ope^{\opi(k z - \omega t)} + O[(\deltaRn)^2].
\end{equation}
Similarly, the perturbed velocity and pressure are of the form 
\begin{equation}
  \label{eq:eqb6}
  \begin{bmatrix}
    \urn(r,z,t) \\ \uzn(r,z,t) \\ \pn(r,z,t)
  \end{bmatrix}
  = 
  \begin{bmatrix}
   \myupper{\bar{u}_r}{n} \\ \myupper {\bar{u}_z}{n} \\ \myupper {\bar{p}}{n}
  \end{bmatrix}
 +  \begin{bmatrix}
    \deltaurn(r) \\ \deltauzn(r) \\ \deltapn(r)
  \end{bmatrix}
  \ope^{\opi (kz - \omega t)}.
\end{equation}
Note that the Stokes equations \eqref{eqStokes} and continuity equation \eqref{eqContinuity} imply that
\begin{equation}\label{eq:eqLaplace}
    \frac{\partial^2 \pn}{ \partial r^2} + \frac{1}{r} \frac{\partial \pn}{\partial r} + \frac{\partial^2 \pn}{\partial z^2} = 0.
\end{equation}
Substituting the third row of \eqref{eqb6} into \eqref{eqLaplace}, we obtain an ordinary differential equation for $\deltapn(r)$ 
\begin{equation}
  \label{eq:eqb10}
  \left(\frac{\opd^2}{\opd r^2} + \frac{1}{r} \frac{\opd }{\opd r} - k^2 \right) \deltapn(r)  = 0.
\end{equation}
Clearly, $\deltapn(r)$ satisfies the modified Bessel equation of order zero in terms of $kr$.  Therefore, we have
\begin{equation}  \label{eq:eqb11}
   \deltapn(r) = \cna   K_0(kr) + \cnb  I_0(kr),
\end{equation}
where $K_0(\cdot)$ and $I_0(\cdot)$ are modified Bessel functions of the first and second kind ($K_0$ is exponentially decreasing and singular at origin; $I_0$ is exponentially growing), and $\cna$ and $\cnb$ are constants to be determined. Inserting $\deltapn(r)$ into \eqref{eqStokes} and solving two inhomogeneous differential equations, we obtain the radial component of velocity
\begin{equation}\label{eq:eqb12}
    \deltaurn(r) =  \cna   \frac{r K_0(kr)}{2 \mun} + \cnb  \frac{r I_0(kr)}{2 \mun} + \cnc  \frac{ K_1(kr)}{2 \mun k} + \cnd  \frac{I_1(kr)}{2 \mun k}
\end{equation}
and the axial component of velocity
\begin{multline}  \label{eq:eqb13}
  \deltauzn(r)  =  \cna   \left[\frac{\opi K_0(kr)}{\mun k}  - \frac{\opi r K_1(kr)}{2 \mun} \right] + \cnb \left[ \frac{\opi I_0(kr)}{\mun k}  + \frac{\opi r I_1(kr)}{2 \mun} \right]\\  - \cnc \frac{ \opi K_0(kr)}{2 \mun k}  +  \cnd  \frac{\opi I_0(kr)}{2 \mun k},
\end{multline}
where $ \cnc $ and $ \cnd $ are constants to be determined. Imposing the conditions that the velocity and pressure must be finite at $r=0$ and $r=+ \infty$, we immediately have 
\begin{equation}\label{eq:eqFinite}
  \myupper{c_1}{1} = \myupper{c_3}{1} = \myupper{c_2}{N} = \myupper{c_4}{N} = 0.
\end{equation}

\subsection{Boundary conditions}\label{sec:BoundaryConditions}
In order to determine the  unknown constants $ \myupper{\vec{c}}{n} = (\cna, \cnb, \cnc, \cnd) $ in each layer, we close the system by imposing boundary conditions at each interface. Let $\nn$ be the unit outward normal vector of interface $r = \zetan(z,t)$ and $\tn$ be the unit tangent vector. Formulae for $\nn$ and $\tn$ are given by equations \eqref{eqVectornj} and \eqref{eqVectortj} of Appendix \ref{AppendixCurvature}. First, the normal component of the velocity is continuous at the interface, since there is no mass transfer across the interface, and so 
\begin{equation}\label{eq:eqFullNSnormal}
  \un|_{r = \zetan} \cdot \nn = \unpa|_{r = \zetan} \cdot \nn.
\end{equation}
For the at-rest steady state \eqref{eqb4N} and \eqref{eqb4}, this condition is equivalent (to first order) to the continuity of radial velocity:
\begin{equation}\label{eq:eqContRadial}
  \urn(r,z,t)|_{r = \zetan} = \myupper {u_r}{n+1}(r,z,t)|_{r = \zetan}.
\end{equation}
Second, the no-slip boundary condition implies that the tangential component of the velocity is continuous at the interface:
\begin{equation}\label{eq:eqFullNStangential}
  \un|_{r = \zetan} \cdot \tn = \unpa|_{r = \zetan} \cdot \tn.
\end{equation}
(The generalization to inviscid fluids, where no-slip boundary conditions are not applied, is considered in Appendix~\ref{SecFullNS}.)
For the at-rest steady state \eqref{eqb4N} and \eqref{eqb4}, this is equivalent (to first order) to the continuity of axial velocity:
\begin{equation}\label{eq:eqContAxial}
  \uzn(r,z,t)|_{r = \zetan} = \myupper {u_z}{n+1}(r,z,t)|_{r = \zetan}.
\end{equation}
Third, the tangential stress of the fluid is continuous at the interface. The stress tensor of the $n$-th fluid in cylindrical coordinates for axisymmetric flow~\mycite{Kundu2007} can be expressed as 
\begin{equation}
  \label{eq:eqstresstensor}
 \taun =  
   \begin{bmatrix} -\pn + 2\mun \frac{\partial \urn }{\partial r}&
     \mun \left( \frac{\partial \urn}{\partial z} + \frac{\partial \uzn}{\partial r}\right) \\
     \mun\left( \frac{\partial \urn}{\partial z} + \frac{\partial \uzn}{\partial r}\right) & 
                    -\pn + 2\mun \frac{\partial \uzn}{\partial z}
     \end{bmatrix}.
\end{equation}
The continuity of the tangential stress at the interface implies that
\begin{equation}\label{eq:eqtangential}
    \nn \cdot \taun|_{r=\zetan} \cdot \tn  =  \nn \cdot \taunpa|_{r=\zetan} \cdot \tn,
\end{equation}
The leading term of \eqref{eqtangential} leads to 
\begin{equation}  \label{eq:eqtangential2}
    \mun \left( \frac{\partial \urn }{\partial z} + \frac{\partial \uzn}{\partial r}\right)\vert_{r= \zetan }  =  \munpa \left( \frac{\partial u_r^{n+1}}{\partial z} + \frac{\partial u_z^{n+1}}{\partial r}\right)\vert_{r= \zetan }.
\end{equation}
Fourth, the jump of the normal stress across the interface must be balanced by the surface-tension force per unit area. The equation for normal stress balance at the interface is 
\begin{equation}\label{eq:eqnormal}
     \nn \cdot (\taunpa -\taun)|_{r=\zetan} \cdot \nn  = \gamman  \kappan, 
\end{equation}
where $\kappan(r,z,t)$ is the mean curvature of the interface. The curvature can be calculated directly from the unit outward normal vector of the interface by $\kappan = \vecnabla \cdot \nn$ (see Appendix \ref{AppendixCurvature}). Substituting \eqref{eqstresstensor}, \eqref{eqVectornj}, and \eqref{eqCurvature2} into \eqref{eqnormal}, we have the following equation (accurate to first order in $\deltaRn$) for the dynamic boundary condition:
\begin{multline}\label{eq:eqnormal2}
    \left[ -\myupper{p}{n+1} + 2\munpa  \frac{\partial \myupper{u}{n+1}}{\partial r} -  \left(-\pn + 2\mun \frac{\partial \urn}{\partial r} \right) \right] \vert_{ r= \zetan } \\ = \gamman \left[ \frac{1}{ \Rn } + \frac{ \deltaurn( \Rn )}{-\opi \omega}  \left(k^2 - \frac{1}{( \Rn )^2} \right) \ope^{\opi (kz - \omega t)} \right].
\end{multline}
\subsection{Dispersion relation}\label{sec:DispersionRelation}
Substituting \eqref{eqb6} and \eqref{eqb11}--\eqref{eqb13} into the boundary conditions [\eqref{eqContRadial}, \eqref{eqContAxial}, \eqref{eqtangential2} and \eqref{eqnormal2}] and keeping the leading terms, we obtain a linear system in terms of the unknown constants $ \myupper {\vec{c}}{n} = (\cna, \cnb, \cnc, \cnd)$. After some algebraic manipulation, these equations can be put into a matrix form:
\begin{equation}\label{eq:eqMatrix1}
   \left(\matAnn + \frac{1}{-\opi \omega} \matBn \right) \myupper {\vec{c}}{n} - \matAnnpa \myupper{\vec{c}}{n+1} = 0,
\end{equation}
where $ \matAnnp $ and $\matBn$ are $4 \times 4$ matrices given below.
\begin{equation}\label{eq:eqMatrixAn}
   \matAnnp  =
  \begin{bmatrix}
   \frac{k  \Rn \Kzn}{\munp} & \frac{k \Rn \Izn}{\munp} &  \frac{\Kan}{\munp} &  \frac{\Ian}{\munp} \\
  \frac{2 \Kzn - k  \Rn  \Kan}{\munp} &  \frac{2 \Izn + k  \Rn  \Ian}{\munp} & -  \frac{\Kzn}{\munp} &  \frac{\Izn}{\munp}\\
   k \Rn  \Kzn - \Kan &  k \Rn  \Izn + \Ian &  \Kan &  \Ian\\
    k \Rn  \Kan & -k \Rn \Ian  &
      \Kzn + \frac{\Kan}{k \Rn }  &
      -\Izn + \frac{\Ian}{k \Rn } 
  \end{bmatrix},
\end{equation}
\begin{equation}\label{eq:eqMatrixBn}
   \matBn  =  \frac{-\gamman k ( 1 - \frac{1}{ (k \Rn)^2})}{ 2 \mun }
  \begin{bmatrix}
   0 &0 &  0 &  0 \\
   0 &0 &  0 &  0 \\
   0 &0 &  0 &  0 \\
   k \Rn  \Kzn &  k \Rn \Izn &  \Kan &  \Ian
  \end{bmatrix},
\end{equation}
Here, $\Kzn,  \Kan, \Izn $ and $\Ian$ denote the corresponding modified Bessel functions evaluated at $k \Rn $.
Combining the boundary conditions from all $N-1$ interfaces, we have the matrix equation
\begin{equation}\label{eq:eqMxi}
  \MN \boldsymbol {\xi} = 0  
\end{equation}
 for the undetermined constants $ \boldsymbol {\xi}  = ( \myupper{c_2}{1}, \myupper{c_4}{1}, \myupper{\vec{c}}{2}, \myupper{\vec{c}}{3}, \cdots, \myupper{\vec{c}}{N-1}, \myupper{c_1}{N}, \myupper{c_3}{N} )$, with
\begin{multline}\label{eq:eqMatrixM}
 \MN = \MaN + \frac{1}{-\opi \omega}\MbN \\
 =\begin{bmatrix}  
  \myupper {\tilde{\matrx{A}}}{1,1} & -\myupper{\matrx{A}}{1,2} &  &   &  &   \\
     & \myupper{\matrx{A}}{2,2} & -\myupper{\matrx{A}}{2,3} &  & & \\    
  &  & \ddots &  \ddots &  &  \\
  &  &  & \myupper{\matrx{A}}{N-2,N-2} & -\myupper{\matrx{A}}{N-2,N-1} &  \\
  & & & & \myupper{\matrx{A}}{N-1,N-1} & -\myupper{\tilde{\matrx{A}}}{N-1,N}
  \end{bmatrix} \\
+ \frac{1}{ -\opi \omega}\begin{bmatrix}  
  \myupper {\tilde{\matrx{B}}}{1} & 0 &  &   &  &   \\
     & \myupper{\matrx{B}}{2} & 0 &  & & \\    
  &  & \ddots & \ddots & &  \\
  &  &  & \myupper{\matrx{B}}{N-2} & 0 &  \\
  & & & & \myupper{\matrx{B}}{N-1} & 0
  \end{bmatrix} ,
\end{multline}
where $\myupper {\tilde{\matrx{A}}}{1,1}$ and $\myupper {\tilde{\matrx{B}}}{1}$ are the second and fourth columns of $\myupper{\matrx{A}}{1,1}$ and $\myupper{\matrx{B}}{1}$, and $\myupper{\tilde{\matrx{A}}}{N-1,N}$ is the first and third columns of $\myupper{\matrx{A}}{N-1,N}$. To obtain a non-trivial solution of equation \eqref{eqMxi}, the coefficient matrix $\MN$ must be singular, namely
\begin{equation} \label{eq:eqDeterminant}
 \det (\MN) = 0.
\end{equation}
Since $\matBn$ is zero except in its fourth row, $\omega$ only occurs in the $4$th, $8$th, $\cdots$, $4(N-1)$-th rows of $\MN$. The Leibniz formula implies that equation \eqref{eqDeterminant} is a polynomial in $1/\omega$ with degree $N-1$. Therefore, we could obtain the dispersion relation $ \omega = \omega(k)$ by solving the polynomial equation \eqref{eqDeterminant}. Instead, to counteract roundoff-error problems, we solve the corresponding generalized eigenvalue problem as described in \mysection\ref{SecMatrixPencil}.
\subsection{Eigen-amplitude and maximum growth rate}\label{SubSecEigenAmp}
The $N-1$ roots of \eqref{eqDeterminant} are denoted by $\omega_j(k)$, where $j=1,2,\cdots,N-1$. Since $\MN$ is singular when $\omega = \omega_j$, $\boldsymbol \xi$ can be determined by equation \eqref{eqMxi} up to a proportionality constant. Therefore, for each $\omega_j(k)$,  the corresponding perturbed amplitude  $\deltaRnj$ on the $n$-th interface can also be determined up to a proportionality constant by equations \eqref{eqb12} and \eqref{eqkinematic}. Let us call the vector $\bolddeltaRj = (\myupper{\delta R_j}{1}, \myupper{\delta R_j}{2}, \cdots,\myupper{\delta R_j}{n}, \cdots, \myupper{\delta R_j}{N-1})$, where we normalize $\Vert \bolddeltaRj \Vert_{1}=1$, the ``eigen-amplitudes'' of frequency $\omega_j$. Any arbitrary initial perturbation amplitudes $\vec{A} = (\myupper{A}{1},\myupper{A}{2},\cdots,\myupper{A}{N-1})$ can be decomposed into a linear combination of eigen-amplitudes, namely $\vec{A} = \sum_{j=1}^{N-1} a_j \bolddeltaRj$ for some $a_j$. Since the whole coupled system is linear, the small initial perturbation $\vec{A} \ope^{\opi k z}$ will grow as $\sum_{j=1}^{N-1} a_j \bolddeltaRj \ope^{\opi [kz - \omega_j(k)t]}$.

The growth rate for a single mode is $\sigma_j(k) = \Im \omega_j(k)$ since our time dependence is $\ope^{-\opi \omega_j t}$. If $\sigma_j > 0$, then the mode is unstable. As described above, for an $N$-layer system, we have $N-1$ different growth rates for a single $k$, and we denote the largest growth rate by $\sigmamax (k) = \max_j [\sigma_j(k)]$. Moreover, the maximum growth rate over all $k$ is denoted by $\sigmamax = \max_k [\sigmamax(k)] = \max_{j,k}[\sigma_j(k)]$, and the corresponding eigen-amplitudes of this most-unstable mode are denoted by $\bolddeltaRmax$; $\kmax$ denotes the corresponding wavenumber.

\section{Generalized eigenvalue problem}
\label{SecMatrixPencil}
It is well known that finding the roots of a polynomial via its coefficients is badly ill-conditioned~\mycite{Trefethen1997}. Correspondingly, we find that solving the determinant equation \eqref{eqDeterminant} directly, by treating it as a polynomial, is highly susceptible to roundoff errors when $N$ is not small. In particular, it is tempting to use the block structure of $\MN$ to reduce \eqref{eqDeterminant} to a $4 \times 4$ determinant problem via a recurrence. However, the entries of this $4 \times 4$ matrix are high-degree polynomials in $\omega$ whose coefficients thereby introduce roundoff ill-conditioning. Instead, we can turn this ill-conditioned root-finding problem into a generalized eigenvalue problem by exploiting the matrix-pencil structure of $\MN$:
\begin{equation}
  \label{eq:eqMatrixPencil}
  \MN \boldsymbol {\xi} = 0 \Leftrightarrow \MbN(k) \boldsymbol {\xi} =  \opi \omega \MaN(k) \boldsymbol {\xi}.
\end{equation}
Thus, finding the dispersion relation $\omega(k)$ turns out to be a generalized eigenvalue problem with matrices $(\MaN, \MbN )$. Since $\MaN$ is non-singular, this (regular) generalized eigenvalue problem is typically well-conditioned~\mycite{Demmel1987} and can be solved via available numerical methods~\mycite{Anderson1999}. In principle, further efficiency gains could be obtained by exploiting the sparsity of this pencil, but dense solvers are easily fast enough for $N$ up to hundreds.
\section{Validation of our formulation}
\label{SecValidation}
As a validation check, our $N$-layer results can be checked against known analytical results in various special cases. We can also compare to previous finite-element calculations~\mycite{Deng2011}.
\subsection{Tomotika's case: $N=2$}\label{SecTomotikaCase}
\citenameasnoun{Tomotika1935} discussed the instability of one viscous cylindrical thread surrounded by another viscous fluid, which is equivalent to our model with $N=2$. It is easy to verify that $\det({\myupper {\matrx{M}}{2}}) = 0$, where $\myupper {\matrx{M}}{2} = [\myupper {\tilde{\matrx{A}}}{1} + \frac{1}{-\opi \omega} \myupper {\tilde{\matrx{B}}}{1}, \myupper {\tilde{\matrx{D}}}{1}]$, gives the same determinant equation as (34) in \citeasnoun{Tomotika1935}. 

In contrast with the Stokes-equations approach, Tomotika began with the full Navier--Stokes equations, treating the densities of the inner fluid $\rho'$ and the outer fluid $\rho$ as small parameters and taking a limit to reach the Stokes regime. However, special procedures must be taken in order to obtain a meaningful determinant equation in this limit, because substituting $\rho'=0$ and $\rho = 0$ directly would result in dependent columns in the determinant. Tomotika proposed a procedure of expanding various functions in ascending powers of $\rho$ and $\rho'$, subtracting the leading terms in dependent columns, dividing a quantity proportional to $\rho \rho'$, and finally taking a limit of $\rho \rightarrow 0$ and  $\rho' \rightarrow 0$. We generalize this idea to the $N$-shell problem in Appendix~\ref{SecFullNS}, but such procedures are unnecessary if the Stokes equations are used from the beginning.

\subsection{$N =3$ with equal viscosities $\myupper {\mu}{1} = \myupper {\mu}{2} = \myupper {\mu}{3}$}
This equal-viscosity case was first discussed by \citenameasnoun{Stone1996}. Putting this special case $\myupper {\mu}{1} = \myupper {\mu}{2} = \myupper {\mu}{3}$ into \eqref{eqDeterminant} and solving it with MATLAB's Symbolic Math Toolbox (MuPAD), we obtain the same solution as equation (8) in \citeasnoun{Stone1996}.

\subsection{Navier--Stokes and inviscid cases}
In Appendix~\ref{SecFullNS}, we validate the generalized form of our instability analysis against previous results for inviscid and/or Navier--Stokes problems. For example, we find identical results to \citeasnoun{Chauhan2000} for the $N=3$ case of a two-fluid compound jet surrounded by air (with negligible air density and viscosity).

\subsection{Comparison with numerical experiments}
\begin{figure}
\centering
 \includegraphics[width=0.6\textwidth]{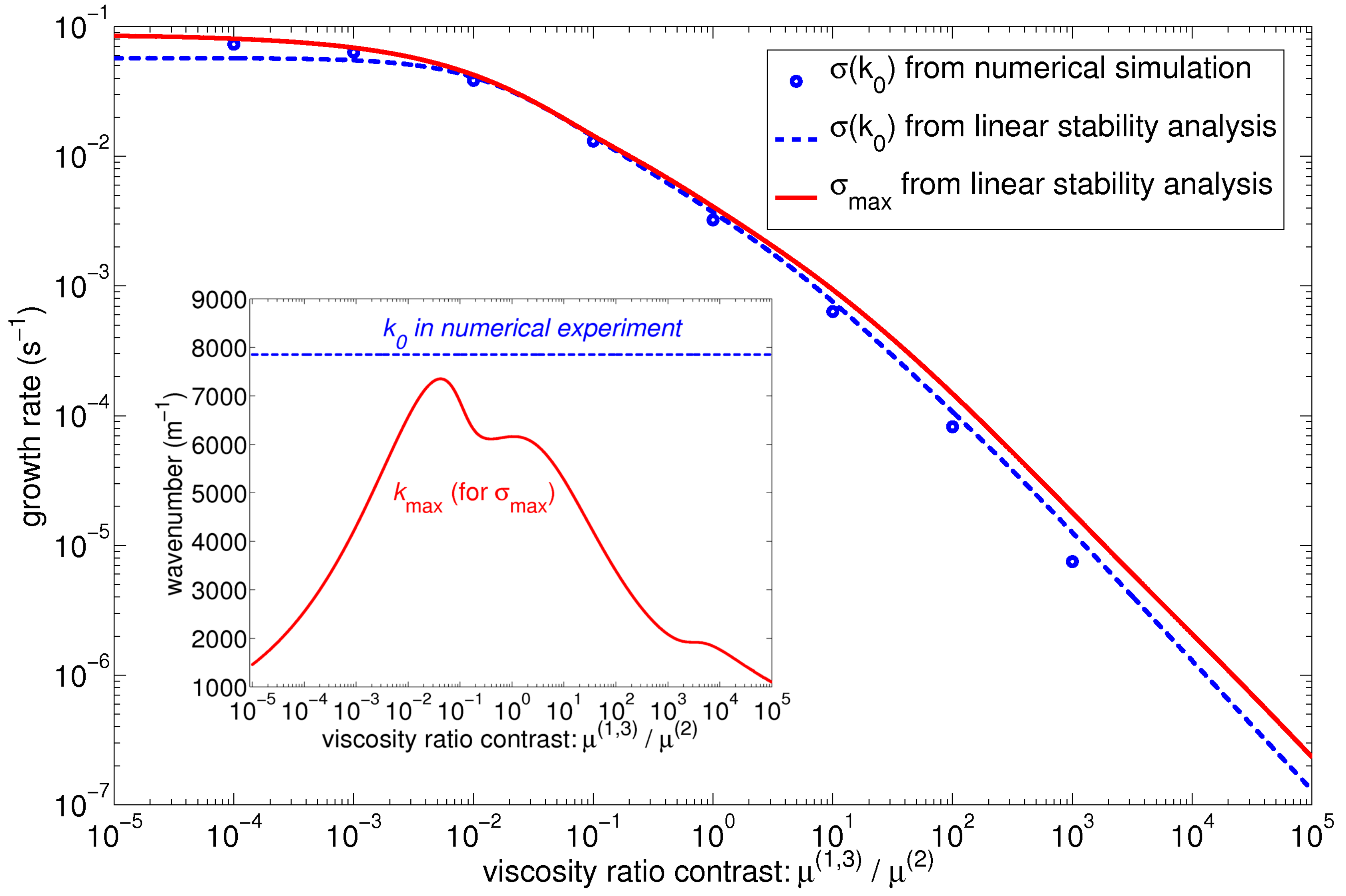}
\setlength{\abovecaptionskip}{0in}
\setlength{\belowcaptionskip}{0in} 
\caption{Comparison between linear stability analysis and numerical experiments [data from~\citeasnoun{Deng2011}] for $N=3$ cylindrical-shell model. The growth rate $\sigmamax(k_0)$ computed by~\citeasnoun{Deng2011} numerically via the finite-element method (blue circles) agrees well with the growth rate predicted by linear stability analysis (dashed blue line), except for small discrepancies in the regime of large viscosity contrast where accurate numerical simulation is difficult. The red line indicates the maximum growth rate $\sigmamax$ obtained by linear stability analysis. In the inset, the red line shows the wavenumbers $\kmax$ for various viscosity-ratio contrasts  and the dashed blue line represents the fixed $k_0$ used in numerical simulations. Model parameters: $\myupper {R}{1} = 60${ $\mu \rm m$}, $\myupper {R}{2}=120 \mu\rm m$, $\myupper{\gamma}{1} = \myupper{\gamma}{2} = 0.6$N/m, $\myupper {\mu}{2} = 10^5$ Pa$\cdot$s, $\myupper {\mu}{1} = \myupper {\mu}{3} =  \eta \myupper {\mu}{2}$   $(\eta = 10^{-4}, 10^{-3}, \cdots, 10^3)$, and $k_0 =  7.9 \times 10^{3} { \rm m}^{-1}$. }
 \label{fig:FigCompareDaosheng}
\end{figure}

\citeasnoun{Deng2011} studied axisymmetric capillary instabilities of the concentric cylindrical shell problem $(N=3)$ for various viscosity contrasts by solving the full Navier--Stokes equation via finite-element methods. [The Stokes equation is a good approximation for their model, in which the Reynolds number is extremely low $( {\rm Re} \approx 10^{-10})$.] In particular, they input a fixed initial perturbation wavenumber $k_0$, evolve the axisymmetric equations, and fit the short-time behaviour to an exponential in order to obtain a growth rate.  With their parameters $\myupper {R}{1} = 60${ $\mu \rm m$}, $\myupper {R}{2}=120 \mu\rm m$, $\myupper{\gamma}{1} = \myupper{\gamma}{2} = 0.6$N/m, $\myupper {\mu}{2} = 10^5$ Pa$\cdot$s, and $\myupper {\mu}{1} = \myupper {\mu}{3} =  \eta \myupper {\mu}{2}$   $(\eta = 10^{-4}, 10^{-3}, \cdots, 10^3)$, we compute the maximum growth rate $\sigmamax$ for each ratio $\eta$ via the equation $\det(\myupper{\matrx{M}}{3}) = 0$. For comparison, we also compute the growth rate $\sigmamax(k_0)$ for their fixed $k_0 = 7.9 \times 10^{3} { \rm m}^{-1} $. [Because numerical noise and boundary artifacts in the simulations will excite unstable modes at $k \ne k_0$, it is possible that $\sigmamax$ and not $\sigmamax(k_0)$ will dominate in the simulations even at short times if the former is much larger.] The inset of \figref{FigCompareDaosheng} plots the wavenumber $k_{\rm max}$ that results in the maximum growth rate versus the viscosity ratio $\eta$ (${\myupper {\mu}{1,3}}/{\myupper {\mu}{2}}$). In \figref{FigCompareDaosheng}, we see that the growth rate obtained by \citeasnoun{Deng2011} (blue circles) agrees well with the growth rate $\sigmamax(k_0)$ predicted by linear stability analysis (blue line) except at large viscosity contrasts ($\eta \gg 1$ or $\eta \ll 1$). These small discrepancies are due to the well-known numerical difficulties in accurately solving a problem with large discontinuities. 
\section{$N=3$ examples}
In this section, we study the three-fluid ($N=3$) problem. Three or more concentric layers are increasingly common in novel fibre-drawing processes~\mycite{Hart2002, Kuriki2004, Pone2006,Abouraddy2007, Deng2008}. By exploring a couple of interesting limiting cases, in terms of shell viscosity and shell thickness, we reveal strong connections between the $N=3$ case and the classic $N=2$ problem.
\subsection{ Case $N = 3$ and ${ \myupper {\mu}{2}}/{\myupper {\mu}{1,3}} \rightarrow 0$: shell viscosity $\ll$ cladding viscosity}
We first consider the limiting case in which the shell viscosity $\myupper{\mu}{2}$ is much smaller than the cladding viscosities $\myupper{\mu}{1}$ and $\myupper{\mu}{3}$. Substituting ${\myupper {\mu}{2}}/{\myupper {\mu}{1}} = 0$ and ${\myupper {\mu}{2}}/{\myupper {\mu}{3}} =0 $ into ${ \myupper{\matrx M}{3}}$, equation \eqref{eqDeterminant} gives simple formulae for the growth rates
\begin{equation}  \label{eq:eqLimit1Sigma1}
  \sigma_1 (k) = \frac{\myupper {\gamma}{1}}{2 \myupper {\mu}{1} \myupper {R}{1}} \frac{ (k \myupper {R}{1})^2 - 1 }{ 1 + (k \myupper {R}{1})^2 - (k \myupper {R}{1})^2 \frac{I_0^2(k\myupper {R}{1})}{I_1^2(k\myupper {R}{1})}}
\end{equation}
and
\begin{equation}  \label{eq:eqLimit1Sigma2}
  \sigma_2 (k) = \frac{\myupper {\gamma}{2}}{2 \myupper {\mu}{3} \myupper {R}{2}} \frac{ 1 - (k \myupper {R}{2})^2}{ 1 + (k \myupper {R}{2})^2 - (k \myupper {R}{2})^2 \frac{K_0^2(k\myupper {R}{2})}{K_1^2(k\myupper {R}{2})}}.
\end{equation}
Note that $\sigma_1(k)$ is independent of  $\myupper {\gamma}{2}$, $\myupper {R}{2}$, and $\myupper {\mu}{3}$, while $\sigma_2(k)$ is independent of $\myupper {\gamma}{1}$, $\myupper {R}{1}$, and $\myupper {\mu}{1}$. In particular, these growth rates are exactly the single-cylinder results predicted by Tomotika's model, as if the inner and outer layers were entirely decoupled. This result is not entirely obvious, because even if the shell viscosity can be neglected, it is still incompressible and hence might be thought to couple the inner and outer interfaces. [\citeasnoun{Deng2011} conjectured a similar decoupling, but only in the form of a dimensional analysis.] 

\subsubsection{Case $N = 3$,  ${\myupper {\mu}{2}}/{\myupper {\mu}{1,3}} \rightarrow 0 \text { and }  \myupper {R}{2} \rightarrow \infty $          $ \Longleftrightarrow   N=2  \text{ and }   {\mu_{\rm out}}/{\mu_{\rm in}} \to 0 $}
In the regime that the shell viscosity $\myupper{\mu}{2}$ is much smaller than the cladding viscosities $\myupper{\mu}{1}$ and $\myupper{\mu}{3}$, we further consider the limit  $\myupper {R}{2} \rightarrow \infty$.  It corresponds to the case $N=2$ with a high viscous fluid embedded in another low-viscosity fluid, which must of course correspond exactly to Tomotika's case. From the asymptotic formulae of modified Bessel functions $K_0(z)$ and $K_1(z)$ for large arguments~\mycite{Abramowitz1992}, we obtain
\begin{equation}\label{eq:eqf1}
   \sigma_2 (k) \approx  - \frac{\myupper {\gamma}{2} |k| }{2 \myupper {\mu}{3}} < 0  \qquad  \text {as} \qquad  \myupper {R}{2} \rightarrow +\infty.  
\end{equation}
Thus, the growth rate of possible unstable modes is given by $\sigma_1(k)$ in equation \eqref{eqLimit1Sigma1}. Tomotika discussed this limiting case ($N=2$) and gave a formula (37)~\mycite{Tomotika1935}, which is exactly \eqref{eqLimit1Sigma1}.

\subsubsection{Case $N = 3 , {\myupper {\mu}{2}}/{\myupper {\mu}{1,3}} \rightarrow 0 \text{ and }  \myupper {R}{1} \rightarrow 0  \Longleftrightarrow   N=2 \text{ and } {\mu_{\rm out}}/{\mu_{\rm in}} \to \infty $ }
The limit $ \myupper {R}{1} \rightarrow 0 $ is equivalent to $N=2$ with a low-viscous fluid embedded in a high-viscosity fluid. For this case, it is easy to check that $\sigma_2(k)$ in equation \eqref{eqLimit1Sigma2} agrees with formula (36) in \citeasnoun{Tomotika1935}. However, we still have another unstable mode with a growth rate $\sigma_1(k)$. Using the asymptotic formulae for $I_0(z)$ and $I_1(z)$ with small arguments~\mycite{Abramowitz1992}, we obtain
\begin{equation}\label{eq:eqf2}
   \sigma_1 (k) \sim   \frac{\myupper {\gamma}{1}  }{6 \myupper {\mu}{1} \myupper {R}{1}}  \qquad  \text{ as} \qquad  \myupper {R}{1} \rightarrow 0.  
\end{equation}
This extra unstable mode $\sigma_1(k)$ results from the instability of a viscous cylinder  with infinitesimally small radius $\myupper {R}{1}$. In other words, \eqref{eqf2} is the growth rate of a viscous cylinder in the air with a tiny but nonzero radius, which is also given by equation (35) of \citeasnoun{Rayleigh1892}. 

\subsection{ Thin shell case: $\myupper {R}{2} =  \myupper {R}{1}(1 + \varepsilon)$, $\varepsilon \to 0$} \label{SubSecThinShell}
\begin{figure}
  \centering
  \subfloat[]{\label{fig:FigThinShellGeo1}\includegraphics[width=0.3\textwidth]{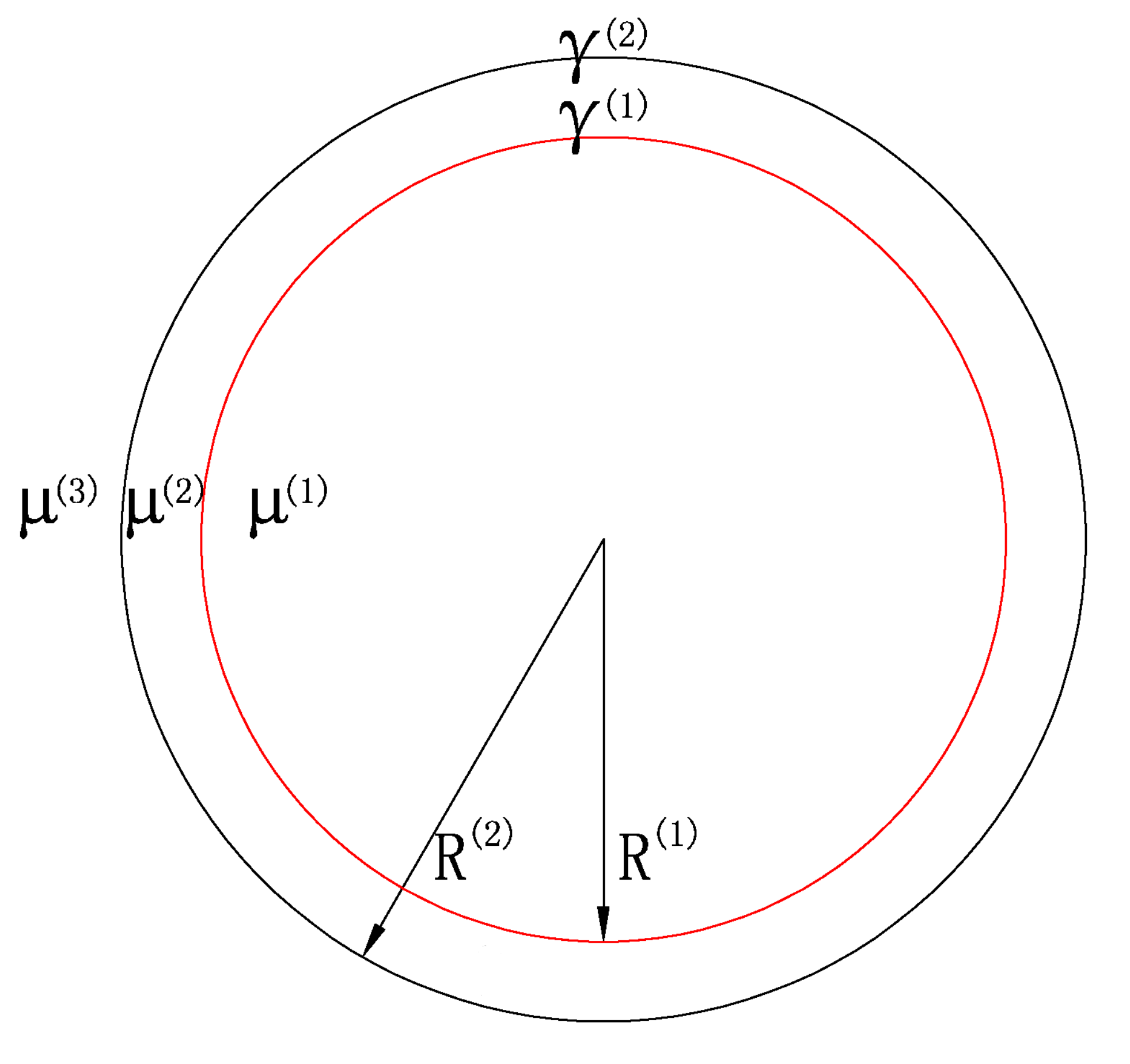}}   \hspace{0.2\textwidth}
  \subfloat[]{\label{fig:FigThinShellGeo2}\includegraphics[width=0.3\textwidth]{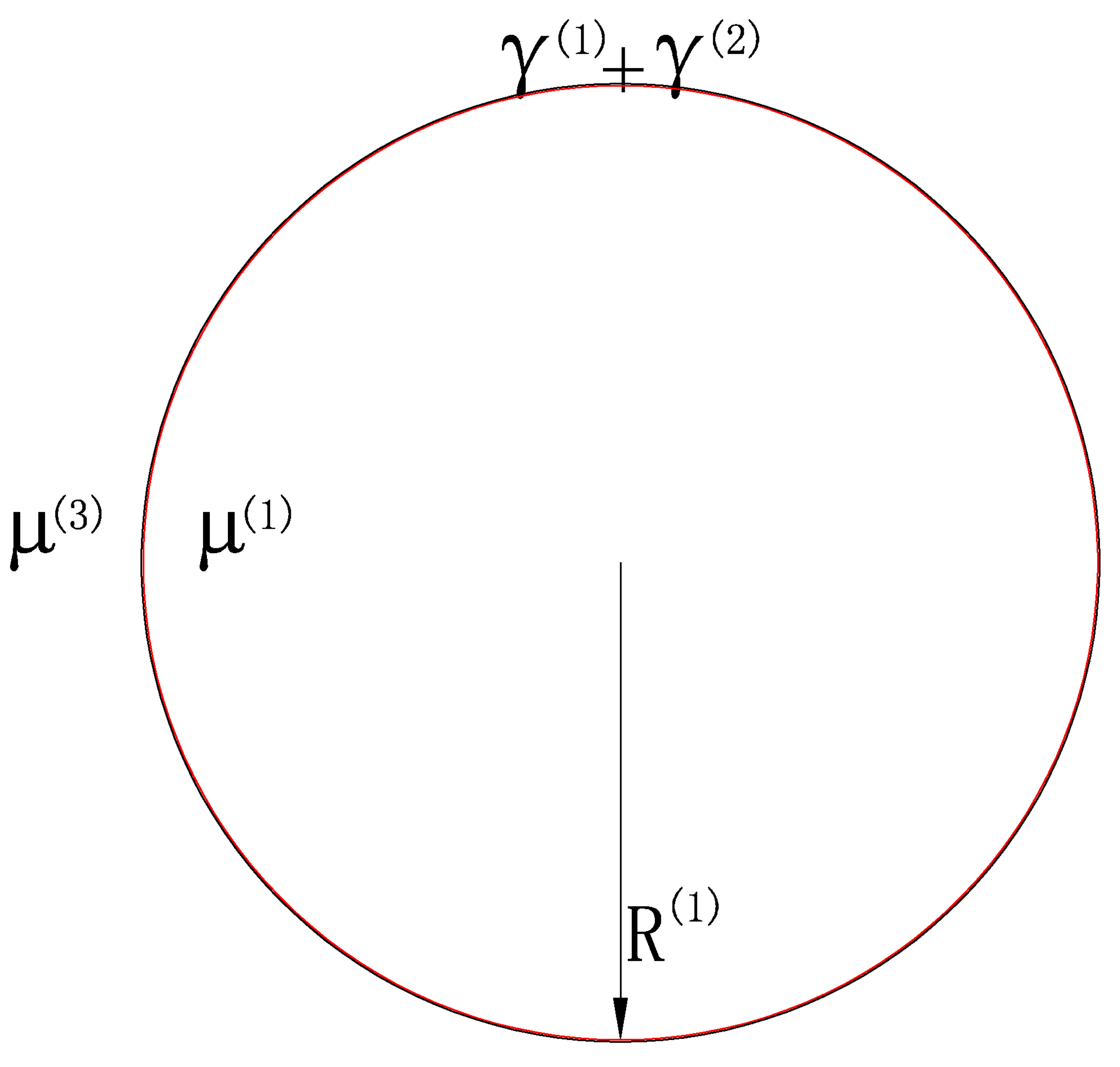}}
  \caption{ (a) Sketch of a very thin shell in a three-layer structure with radius $\myupper {R}{2} =  \myupper {R}{1} (1 + \varepsilon)$, surface-tension coefficients $\myupper{\gamma}{1}$ and $\myupper{\gamma}{2}$. (b) In the limit of infinitesimal $\varepsilon$, we obtain an equivalent $N=2$ geometry with a modified surface-tension coefficient $\myupper {\gamma}{1} + \myupper {\gamma}{2}$. }
  \label{fig:FigThinShellGeoBoth}
\end{figure}

Next, we study a three-layer structure with a very thin middle shell; that is, $\myupper {R}{2} =  \myupper {R}{1} (1 + \varepsilon)$ with $\varepsilon \to 0$.  A sketch of such a geometry is given in  \figref{FigThinShellGeo1}.  This is motivated by a number of experimental drawn-fibre devices, which use very thin (sub-micron) layers in shells hundreds of microns in diameter in order to exploit optical interference effects~\mycite{Hart2002, Kuriki2004,Pone2006}.

\begin{figure}
  \centering
 \subfloat[In-phase mode $\sigma^{+}$]{\label{fig:FigInPhaseMode}\includegraphics[width=0.47\textwidth]{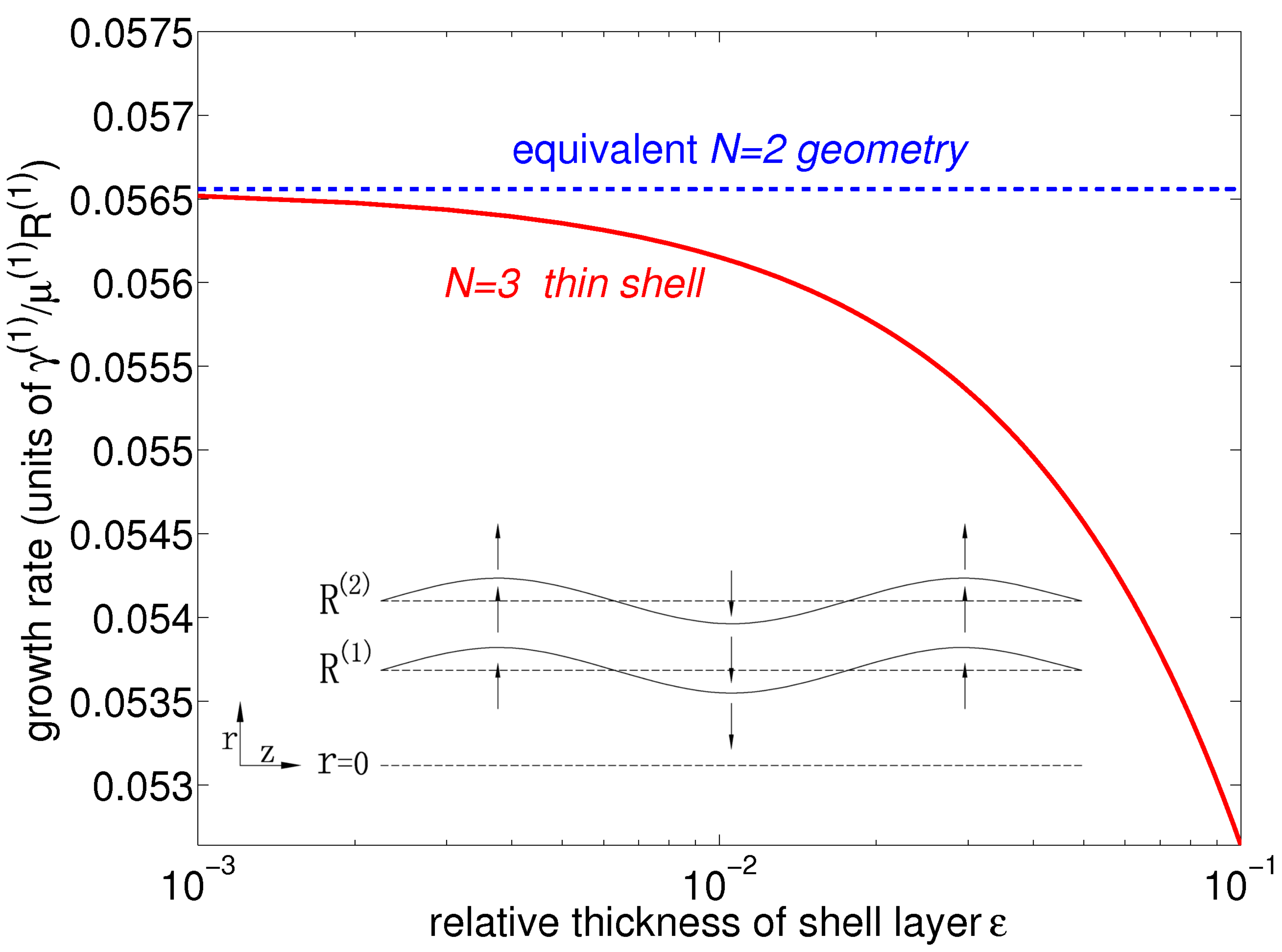}}
  \subfloat[Out-of-phase mode $\sigma^{-}$]{\label{fig:FigOutOfPhaseMode}\includegraphics[width=0.47\textwidth]{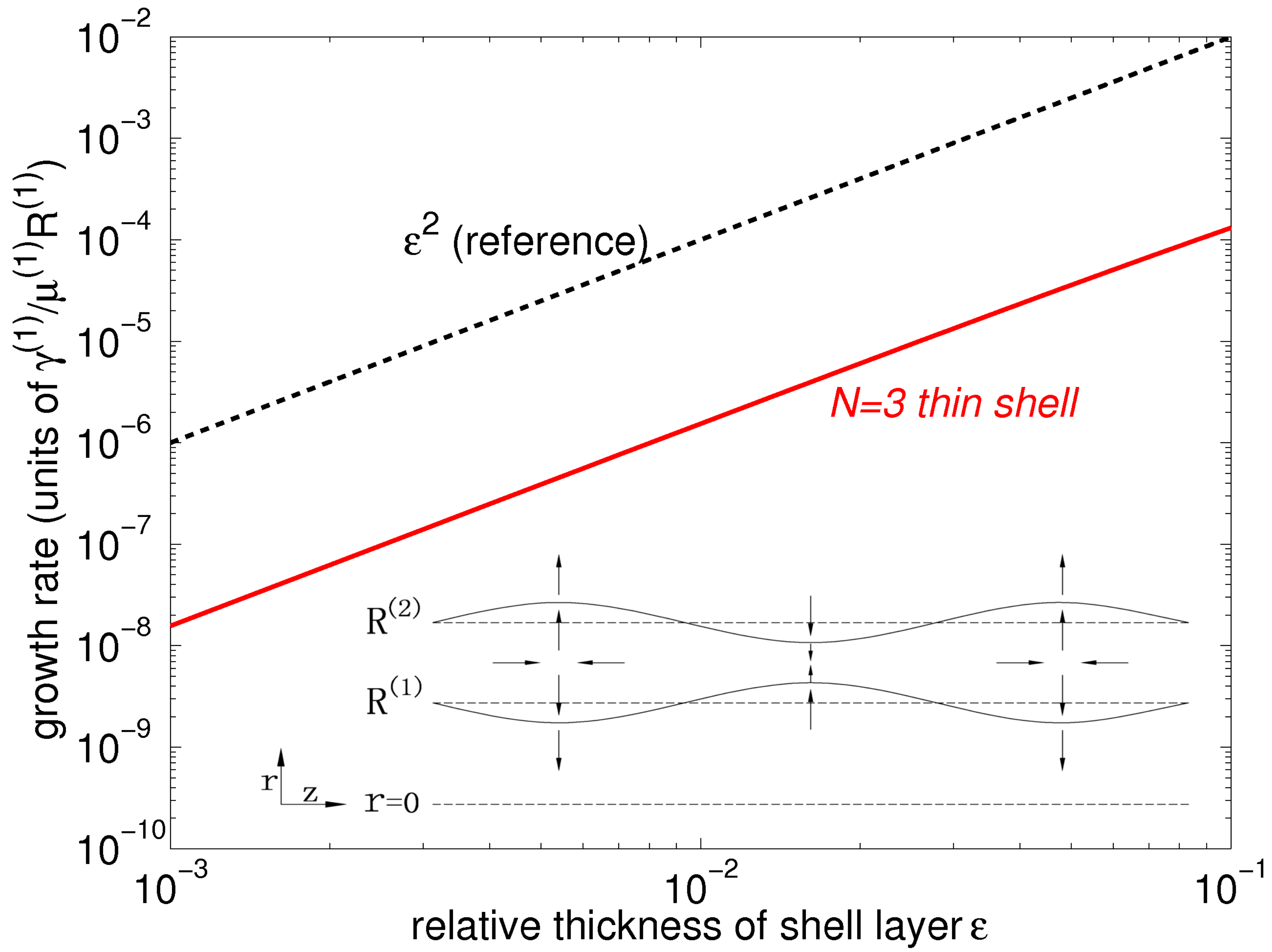}}
  \caption{Two modes $\sigma^{+}$ and $\sigma^{-}$ for thin-shell layer geometry, with $\myupper{R}{1} = 1$, $\myupper {R}{2} =  \myupper {R}{1}(1 + \varepsilon), \myupper{\gamma}{1}=1, \myupper{\gamma}{2}=2, \myupper{\mu}{1} = 1, \myupper{\mu}{2}=2$, $\myupper{\mu}{3} = 3$ and $k=0.5$. (a) In-phase mode $\sigma^+$ illustrates that the growth rate of the in-phase mode $\sigma^+(k)$ for $N=3$ approaches to the growth rate of $N=2$ structure with the summed surface-tension coefficients as $\varepsilon \to 0$. (b) Out-of-phase mode $\sigma^{-}$ demonstrates that the out-of-phase growth rate  $\sigma^{-}(k)$ decreases like $\varepsilon^2$ as $\varepsilon \to 0$. }
  \label{fig:FigTwoModes}
\end{figure}

Considering $\varepsilon$ as a small parameter, we expand the determinant equation \eqref{eqDeterminant} in powers of $\varepsilon$. For a given wavenumber $k$, the two roots of this equation are $\sigma^{+}(k) = \sigma_0(k) + O(\varepsilon)$ and $\sigma^{-}(k)=O(\varepsilon^2)$, where $\sigma_0(k)$ can be computed analytically by dropping the terms of order $O(\varepsilon^2)$ in the determinant equation \eqref{eqDeterminant}. After some algebraic manipulation, we find that $\sigma_0(k)$ actually is the solution for the $N=2$ structure (i.e., ignoring the thin shell) with a modified surface-tension coefficient $\myupper {\gamma}{1} + \myupper {\gamma}{2}$. It is also interesting to consider a limit in which $\myupper{\mu}{2}$ grows as $\varepsilon$ shrinks. In this case, we find the same asymptotic results as long as $\myupper{\mu}{2}/\myupper{\mu}{1,3}$ grows more slowly than $1/\varepsilon$. Conversely, if it grows faster than $1/\varepsilon$, then the thin-shell fluid acts like a ``hard wall'' and all growth rates vanish. Instead of presenting a lengthy expression for $\sigma^{+}(k)$, we demonstrate a numerical verification in  \figref{FigTwoModes}. As indicated in \figref{FigInPhaseMode}, the growth rate $\sigma^+(k=0.5)$ for $N=3$ approaches the the growth rate of $N=2$ with the summed surface-tension coefficients as $\varepsilon \to 0$. The  parameters are $\myupper{R}{1} = 1, \myupper{\gamma}{1}=1, \myupper{\gamma}{2}=2, \myupper{\mu}{1} = 1,  \myupper{\mu}{2} = 2 $ and $\myupper{\mu}{3} = 3$.  In \figref{FigOutOfPhaseMode}, we show that the growth rate $\sigma^{-}(k)$ decreases like $\varepsilon^2$ as $\varepsilon \to 0$.

 To better understand these two modes, we consider the eigen-amplitudes at the two interfaces. For the mode with growth rate $\sigma^{+}(k)$, two interfaces are moving exactly in phase.  Since the thickness of this shell is so thin, it is not surprising that one can treat two interfaces as one with a modified surface-tension coefficient $\myupper {\gamma}{1} + \myupper {\gamma}{2}$ for this mode (see  \figref{FigThinShellGeo2} and the inset of  \figref{FigInPhaseMode}). The eigen-amplitude (defined in section \ref{SubSecEigenAmp}) corresponding to this in-phase mode is $(1/2,1/2)$, independent of $\myupper{\gamma}{1}$ and $\myupper{\gamma}{2}$. For the other mode, with growth rate $\sigma^{-}(k)$, the two interfaces are moving out of phase (see the inset of  \figref{FigOutOfPhaseMode}). The eigen-amplitude for this out-of-phase mode is found to be $(-\myupper{\gamma}{2},\myupper{\gamma}{1})/(\myupper{\gamma}{1} +\myupper{\gamma}{2})$, which means that the two interfaces are moving in opposite directions with amplitudes inversely proportional to their surface tensions. Due to the tiny thickness of the shell compared to its radius of curvature, this case approaches the case of a flat sheet, which is known to be always stable~\mycite{Drazin2004}, as can be proved via a surface-energy argument.

A related thin-shell problem was investigated by \citeasnoun{Chauhan2000} for the Navier--Stokes equations with an inviscid (gaseous) outer fluid. Those authors also found that the problem reduced to $N=2$ instabilities (single fluid surrounded by gas) with a summed surface tension.

\section{Effective growth rate and competing  modes} \label{SecCompetingModes}
In previous work on linear stability analysis, most authors identified the maximum $\sigma$ with the dominant breakup process~\mycite{Rayleigh1879, Rayleigh1892, Tomotika1935}. This exclusive emphasis on the maximum $\sigma$ was continued in recent studies of $N=3$ systems \mycite{Stone1996, Chauhan2000}, but here we argue that the breakup process is more complicated for $N>2$. In a multi-layer situation, however, there is a geometric factor that complicates this comparison: not only are there different growth rates $\sigma$, but there are also different length scales $\Rn$ over which breakup occurs. As a result, it is natural to instead compare a breakup time scale given by a distance ($\tilde{R}$) divided by a velocity, where $\tilde{R}$ is some average radius for a given growth mode (weighted by the unit-norm eigen-amplitudes $\myupper{\delta R}{n}$). In our case, we find that a harmonic-mean radius $\tilde{R}$ is convenient, and we define an effective growth rate ($\sim 1/{\rm{breakup \ time}} \sim {\rm velocity}/{\tilde{R}}$) by
\begin{equation}
   \label{eq:eqEffGrowth}
   \sigmaeff_j(k) = \sigma_j(k) \sum_{n=1}^{N-1} \frac{\myupper{\delta R_j}{n}}{\Rn}.
\end{equation}

\begin{figure}
\centering
 \includegraphics[width=0.6\textwidth]{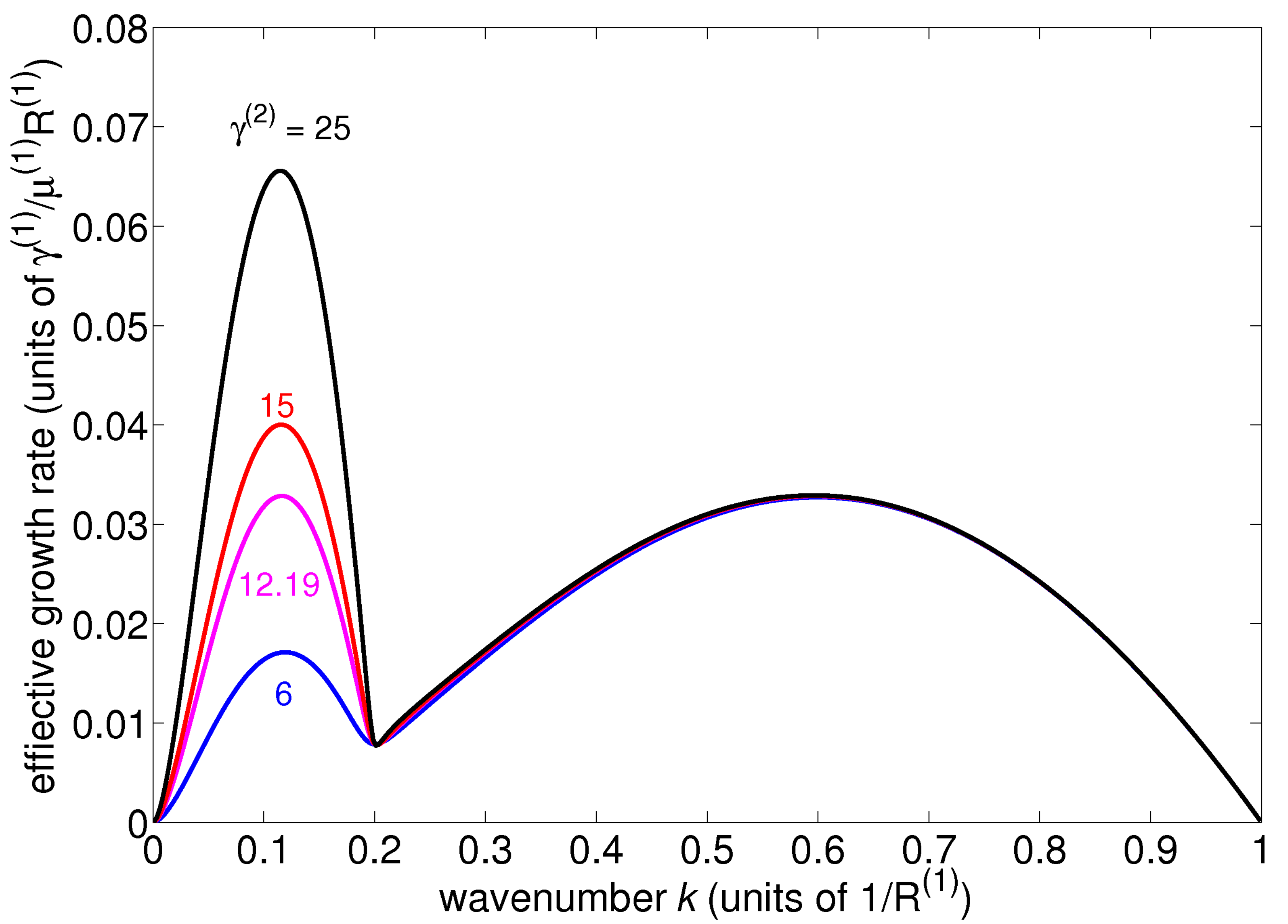}
\setlength{\abovecaptionskip}{0in}
\setlength{\belowcaptionskip}{0in} 
\caption{Maximum effective growth rates versus \ wavenumber k. For a three-layer structure with  $\myupper{R}{1} = 1, \myupper{R}{2} = 5, \myupper{\mu}{1} = \myupper{\mu}{2} = 1$, and $\myupper{\gamma}{1} = 1$, the maximum effective growth rates $\sigmaeffmax(k)$ are plotted for several values of $\myupper{\gamma}{2}$. For $\myupper{\gamma}{2} = 12.19$ (magenta line), there are two equal maximum effective growth rates $\sigmaeffmax(k_1 \approx 0.58) = \sigmaeffmax(k_2 \approx 0.114)$.}
 \label{fig:FigDiffGammaDispersion}
\end{figure}
Now, it is tempting to wonder what happens if two different wavenumbers $k_1$ and $k_2$ have the same maximum effective growth rates, a question that does not seem to have been considered in previous linear stability analyses. Let us consider a particular three-layer structure with $\myupper{R}{1} = 1, \myupper{R}{2} = 5, \myupper{\mu}{1} = \myupper{\mu}{2} = 1$, and $\myupper{\gamma}{1} = 1$. The maximum effective growth rates $\sigmaeffmax(k) =  \max_{j}[\sigmaeff_j(k)]$ versus\ $k$ are plotted in \figref{FigDiffGammaDispersion} for several values of $\myupper{\gamma}{2}$. For example, at $\myupper{\gamma}{2} = 12.19$, we find that $\sigmaeffmax(k_1 \approx 0.58) = \sigmaeffmax(k_2 \approx 0.114)$, so that there are two competing modes at very different length scales $ 2\pi/k_1 = 10.83  $ and $ 2\pi/k_2 = 55.12$. In contrast, for $\myupper{\gamma}{2} = 6$ we see that the short length-scale instability should dominate, while at $\myupper{\gamma}{2} = 25$ the long length scale instability should dominate. 
 
\begin{figure}
  \centering
 \subfloat[]{\label{fig:FigWhiteNoise}\includegraphics[width=0.14\textwidth]{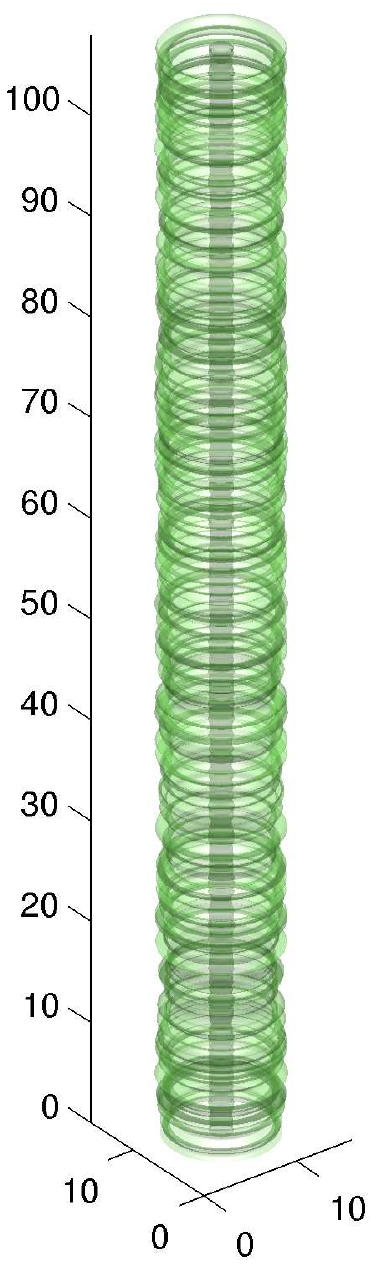}}
\hspace{0.05\textwidth}
 \subfloat[$\myupper{\gamma}{2} = 6$ ]{\label{fig:FigGamma6}\includegraphics[width=0.14\textwidth]{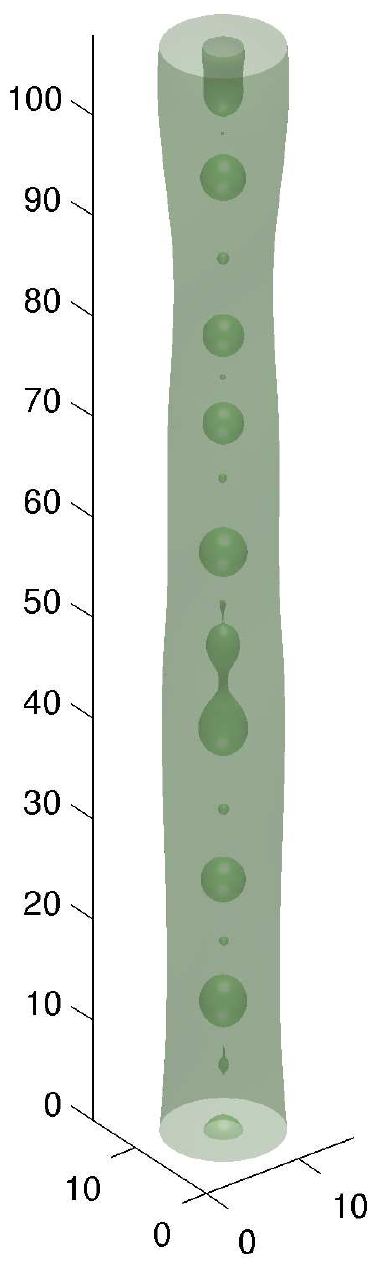}}
\hspace{0.05\textwidth}
  \subfloat[$\myupper{\gamma}{2} = 15$]{\label{fig:FigGamma15}\includegraphics[width=0.14\textwidth]{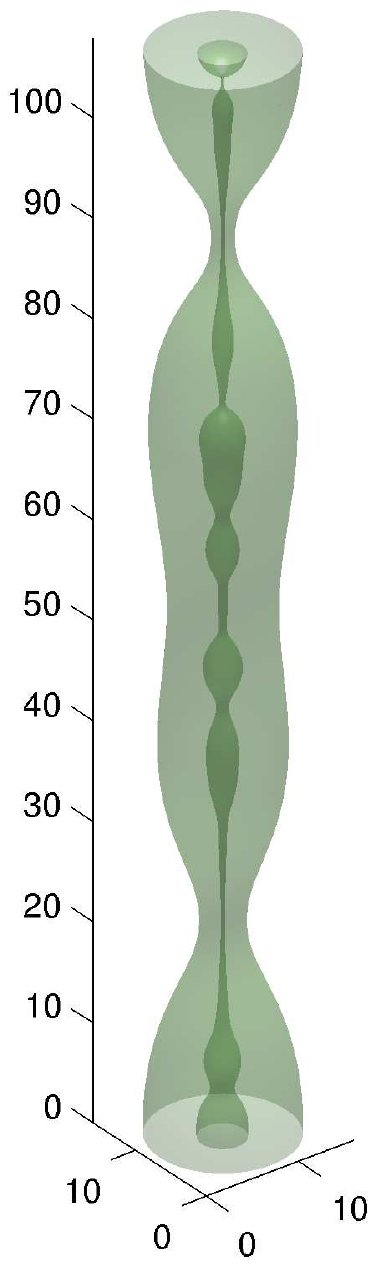}}
\hspace{0.05\textwidth}
 \subfloat[$\myupper{\gamma}{2} = 25$]{\label{fig:FigGamma25}\includegraphics[width=0.135\textwidth]{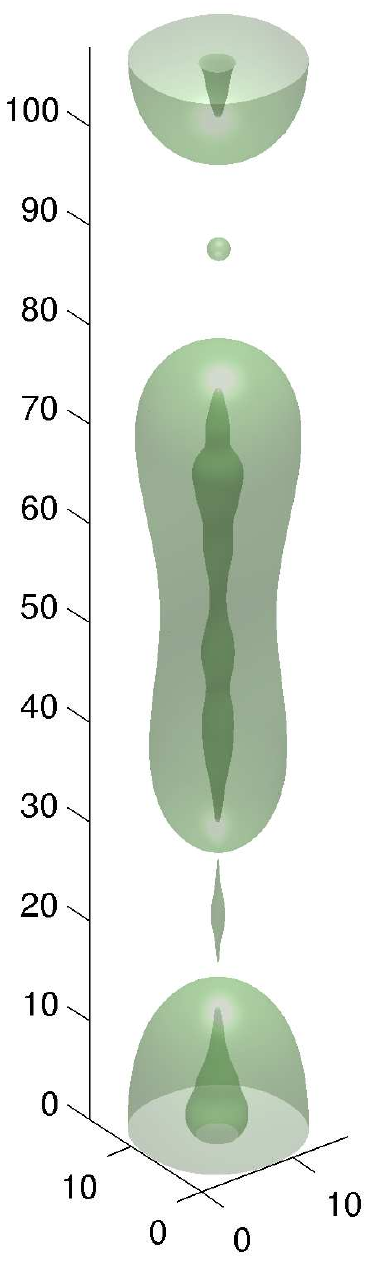}}
  \caption{Numerical Stokes-flow simulations for three-layer systems with different $\myupper{\gamma}{2}$. (a) Initial white-noise perturbations of the interfaces. As predicted by maximum effective growth rates, the systems with $\myupper{\gamma}{2} = 6$ (b) and $\myupper{\gamma}{2} = 25$ (d) exhibit breakup initially via the short- and long-scale modes, respectively (which are dominated by motion of the inner and outer cylinders, respectively). Near-simultaneous breakup occurs for $\myupper{\gamma}{2} = 15$ (c). }
  \label{fig:FigDiffGammaNum}
\end{figure}
To test our predictions, we implemented a full three-dimensional Stokes-flow numerical scheme to simulate the breakup process of this cylindrical-shell system. A brief description of this hybrid scheme,  a combined spectral and level-set method, is given in Appendix \ref{SecNumScheme}. We use initial white-noise perturbations on both interfaces $\myupper{R}{1}$ and $\myupper{R}{2}$ (see \figref{FigWhiteNoise}). The computational box is $16 \times 16 \times 108$ with resolutions $160 \times 160 \times 480$ pixels. As predicted, $\myupper{\gamma}{2} = 6$ and $\myupper{\gamma}{2} = 25$ exhibit breakup initially via the short- and long-scale modes, respectively (which are dominated by motion of the inner and outer cylinders, respectively). It is interesting to estimate the intermediate $\myupper{\gamma}{2}$ where the two breakup processes occur simultaneously. Linear stability analysis predicts $\myupper{\gamma}{2} \approx 12.19$, and indeed we find numerically that near-simultaneous breakup occurs for $\myupper{\gamma}{2} \approx 15$ (see \figref{FigGamma15}). In contrast, simply looking at $\sigmamax$ rather than $\sigmaeffmax$ would lead one to predict that simultaneous breakup occurs at $\myupper{\gamma}{2} \approx 4.15$, in which case all three $\myupper{\gamma}{2}$  values in \figref{FigDiffGammaNum} would have looked like \figref{FigGamma25} (large scale dominating). In the case of near-simultaneous breakup timescales, the dominant breakup process may be strongly influenced by the initial conditions (i.e. the initial amplitudes of the modes), which offers the possibility of sensitive experimental tunability of the breakup process.

\begin{figure}
\centering
 \includegraphics[width=0.9\textwidth]{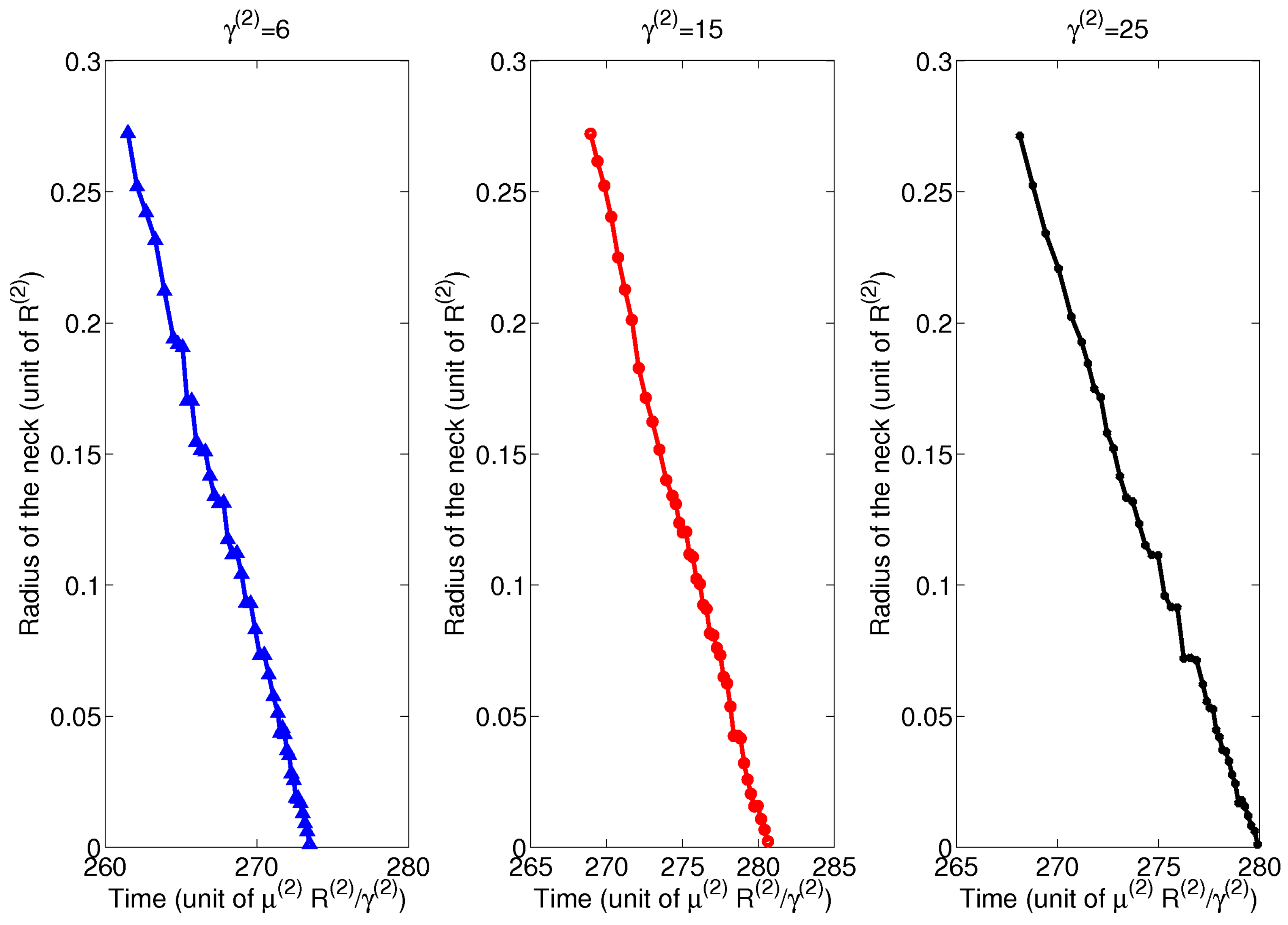}
\caption {Radius of the fluid neck versus time during the final phase of the breakup of the outermost interface, from the 3-dimensional Stokes simulations of \figref{FigDiffGammaNum}. This breakup is asymptotically linear with time, similar to the predictions of the scaling theory for $N=2$ systems \mycite{Lister1998, Cohen1999}.}
 \label{fig:FigNeckRadius}
\end{figure}

The final breakup of the fluid neck is described by a self-similar scaling
theory for the case of a single fluid jet \mycite{Eggers1993}, and so it is interesting to examine numerically to what extent a similar description is
possible for the $N=3$ system. In particular, at the last stage of a single-cylinder breakup process, a singularity develops at the point of breakup which does not possess a characteristic scale, and hence a set of self-similar profiles can be predicted \mycite{Eggers2008}. For both a viscous jet in gas \mycite{Eggers1993} and a viscous thread in another viscous fluid ($N=2$)  \mycite{Lister1998, Cohen1999}, these principles predict that the neck radius $h(t)$ vanishes linearly with time as $h(t) \mu /\gamma \sim (t_0-t)$ where $t_0$ is the breakup time. However, there is no available scaling theory for $N \ge 3 $ systems. Here, we simply use our numerical simulations above to study the rate at which the neck radius vanishes in an $N=3$ system. For all three cases with different $\myupper{\gamma}{2}$, the neck radius of the outer interface vanishes with time in an asymptotically linear fashion as the breakup time is approached (see \figref{FigNeckRadius}). This is not surprising in the $\myupper{\gamma}{2} = 6$ case where the inner surface has already broken up---the breakup of the outer surface reduces to an $N=2$
problem when the neck becomes thin enough---and we find that $h(t) \mu/\myupper{\gamma}{2} \approx 0.024 (t_0 - t) $, in reasonable agreement with the $0.033$ value predicted analytically \mycite{Cohen1999} given the low spatial resolution with which we resolve the breakup singularity ($h/\myupper{R}{2} = 0.1$ corresponds to 5~pixels). Moreover, we find that in this equal-viscosity $N=3$ system, all three $\myupper{\gamma}{2}$ values yield slopes of $h(t)\mu/\myupper{\gamma}{2}$ that are within 10\% of one another, indicating that the inner-surface tension $\myupper{\gamma}{1}=1$  has a relatively small impact on the outer-surface breakup.

\section{$N$-Layer structures}\label{SecNShell}
Since all previous work has studied only $N=2$ or $N=3$, it is interesting to consider the opposite limit of $N \to \infty$. We consider two examples: a repeating structure of two alternating layers, and a structure with continuously varying viscosity, both of which are approached as $N \to \infty$. In fact, concentric-shell structures with dozens of alternating fluid layers have been used experimentally in optical fibres \mycite{Hart2002, Kuriki2004}. However, the motivation of this section is primarily exploratory, rather than engineering---to begin to discover what new phenomena may arise for large $N$.
\subsection{Alternating structure}
\begin{figure}
\centering
 \includegraphics[width=0.7\textwidth]{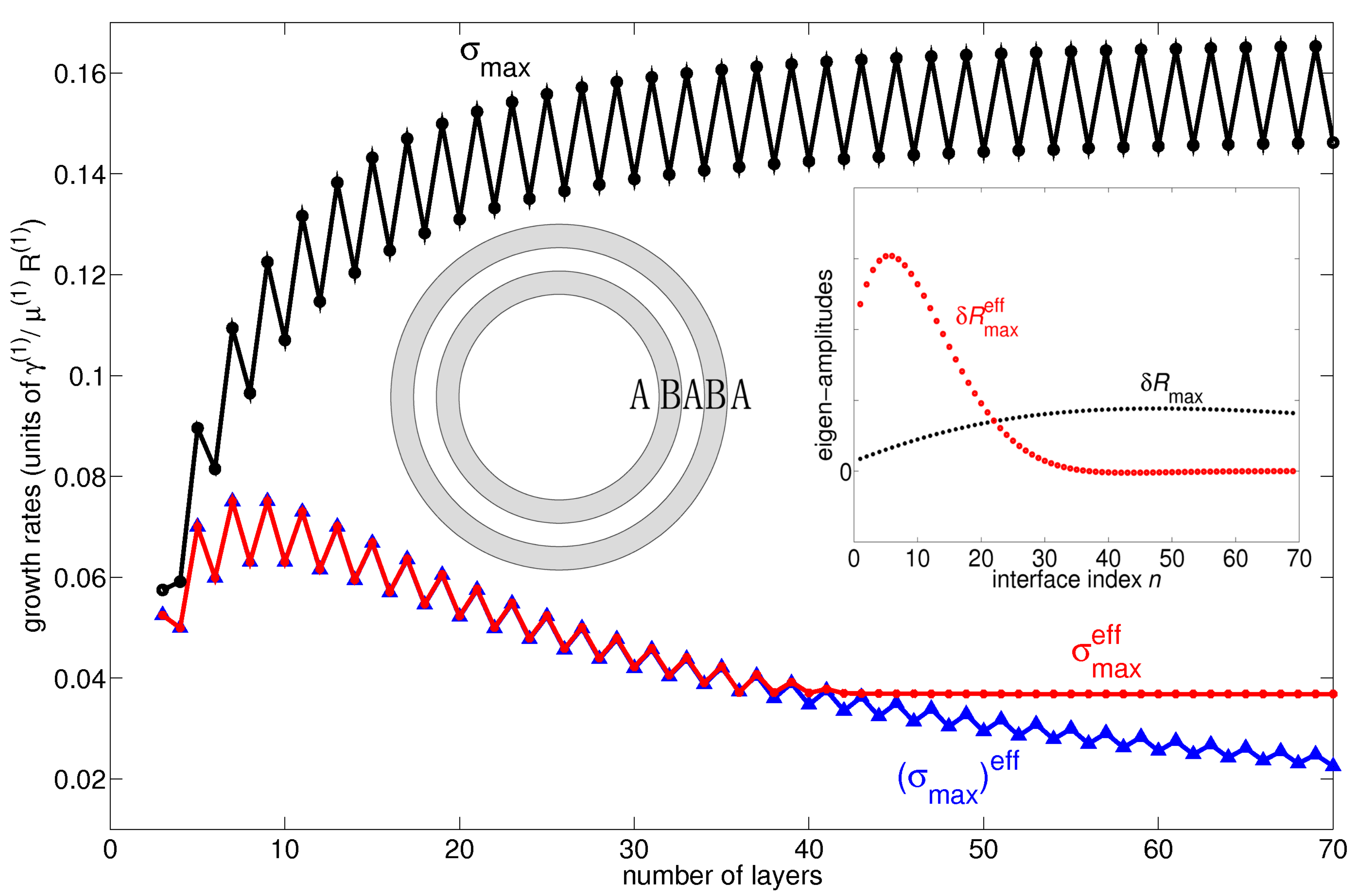} 
\caption{Growth rates of an  $ABABAB\cdots$  alternating structure. Both $\sigmamax$ (black line) and $\sigmaeffmax$ (red line) converge to finite asymptotic values as $N \to \infty$, although in the former case the asymptotic value depends on the parity of $N$. The differences between $\sigmaeffmax$ (red line) and $(\sigmamax)^{\rm eff}$ (blue line)  imply that the modes corresponding to $\sigmamax$ and $\sigmaeffmax$  are not always the same.  The right inset shows the eigen-amplitudes $\bolddeltaRmax$ (black dots) and $\bolddeltaRmaxeff$ (red circles) for $N=70$, corresponding to $\sigmamax$ and $\sigmaeffmax$ respectively. The left inset depicts the structure whose parameters are $\Rn = 1 + 0.2(n-1)$, $\gamman = 1$, and $\mun = 1$ if $n$ is odd or $\mun = 2$ otherwise. }
 \label{fig:FigAlterNLayer}
\end{figure}
First we consider an $ABABAB\cdots$ structure of two alternating, repeating layers $A$ and $B$ as shown schematically in the left inset of  \figref{FigAlterNLayer}. We choose $ \mun = 1$ if $n$ is odd and $\mun = 2$ otherwise. The other parameters are $\Rn = 1 + 0.2(n-1)$ and $\gamman = 1$. 

For this multilayer structure, we find that both the maximum growth rate $\sigmamax$ of the fastest-growing mode and equation {\eqref{eqEffGrowth}}'s maximum effective growth rate $\sigmaeffmax$ (corresponding to the shortest breakup time scale) apparently converge to finite asymptotic values as $N \to \infty$ (\figref{FigAlterNLayer}). (We have checked that the absolute value of the slope of $\sigmamax$ is monotonically decreasing for a broader range $N$ values up to $N=120$, and the slope is $\sim 10^{-5}$ for $N=120$. A rigorous proof of convergence requires a more difficult analysis, however.) The oscillations in  \figref{FigAlterNLayer} are due to the varying viscosity of the ambient fluid, which depends on the parity of $N$. It is interesting to know whether the fastest-growing mode and the mode with maximum effective growth are identical for a given $N$. To see this, we plot the effective growth rate of the fastest-growing mode $(\sigmamax)^{\rm eff}$ versus $N$ and compare it with $\sigmaeffmax$ versus $N$ in \figref{FigAlterNLayer}. Note that $(\sigmamax)^{\rm eff} = \{\max_{j,k} [\sigma_j(k)]\}^{\rm eff}$ and $\sigmaeffmax = \max_{j,k}[\sigma_j^{\rm eff}(k)]$. From \figref{FigAlterNLayer}, we can see that $(\sigmamax)^{\rm eff}$ and $\sigmaeffmax$ are different for large $N$, which implies that the modes corresponding to $\sigmamax$ and $\sigmaeffmax$ are not always the same. 

In the right inset of \figref{FigAlterNLayer}, we plot the eigen-amplitudes $\bolddeltaRmax$ and $\bolddeltaRmaxeff$ for $N=70$, corresponding to $\sigmamax$ and $\sigmaeffmax$ respectively. The mode $\sigmamax$ is mostly motion of outer interfaces, while the mode $\sigmaeffmax$ is mostly motion of inner interfaces. The mostly outer-interface motion for $\sigmamax$ explains why the value of $\sigmamax$ oscillates depending on the ambient fluid. Physically, the association of $\sigmaeffmax$ with the inner interfaces makes sense because, in our definition \eqref{eqEffGrowth} of effective breakup rate, it is easier to break up at smaller radii (a smaller distance to breakup). Alternatively, if we defined ``breakup distance'' in terms of the thickness of individual layers, then $\sigmamax$ would make more sense.

\subsection{$N$-Layer structure for a continuous model}
\begin{figure}
\centering
 \includegraphics[width=0.7\textwidth]{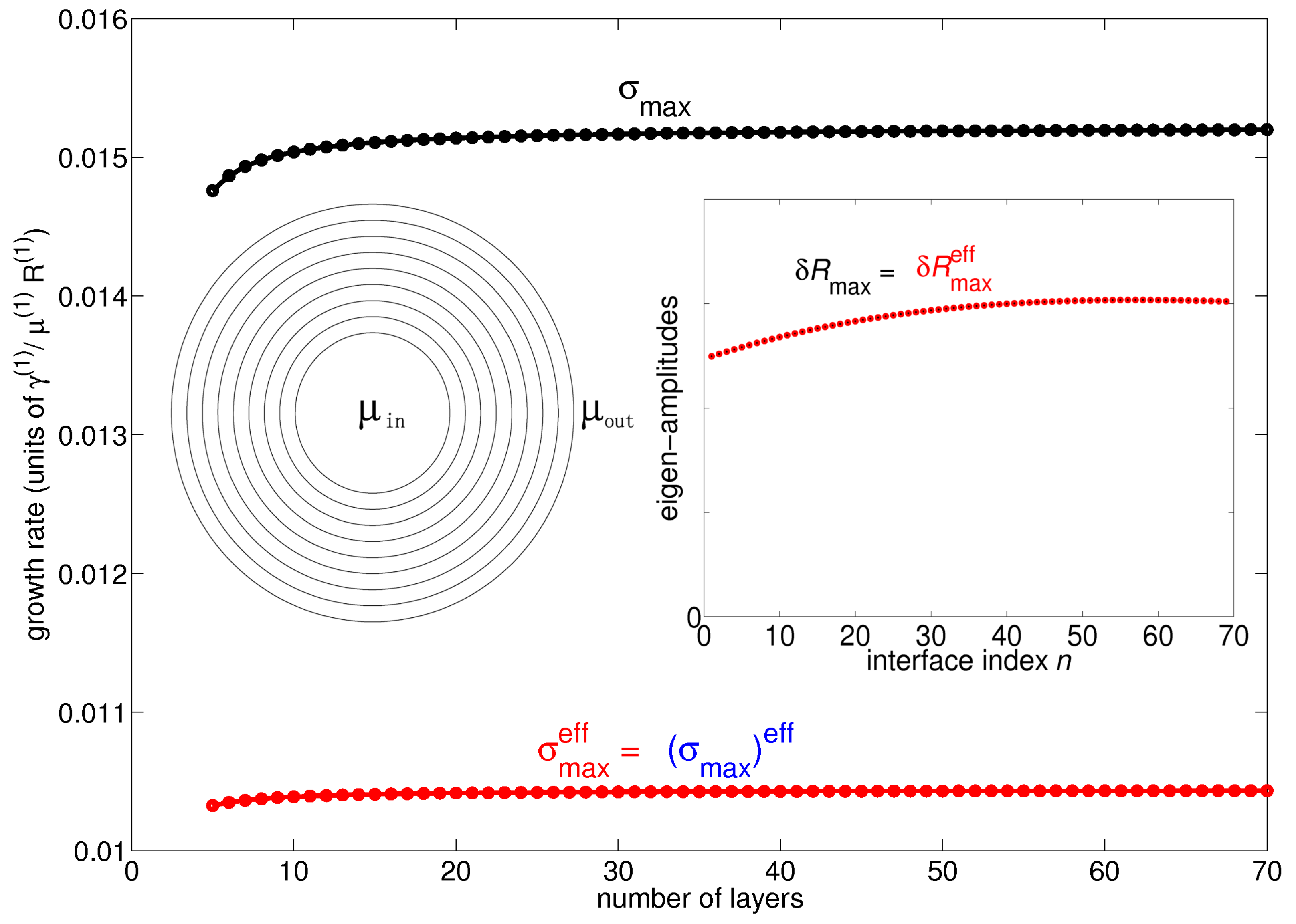}
\caption{Growth rates for a continuous $N$-layer structure.  Both $\sigmamax$ and $\sigmaeffmax$ approach constants as $N \to \infty$. The right inset plots the corresponding eigen-amplitudes $\bolddeltaRmax$ and $\bolddeltaRmaxeff$ (for $N$=70). The left inset sketches the $N$-layer structure: radius $\myupper{R}{1} = \Rin$, $\myupper{R}{N-1} = \Rout$,  $\Rn = \myupper{R}{1} + \frac{\myupper{R}{N-1} - \myupper{R}{1}}{N-2}(n-1)$,  and viscosity $\myupper{\mu}{1} = \muin$, $\myupper{\mu}{N} = \muout$, and $\mun = \myupper{\mu}{1} + \frac{\myupper{\mu}{N} - \myupper{\mu}{1}}{N-1}(n-1)$, approximating a continuously and linearly varying three-layer viscosity. }
 \label{fig:FigContNLayer}
\end{figure}
In this subsection, we build an $N$-layer model to approximate a three-layer structure with a continuous viscosity. The viscosity of intermediate layer $\mumid$ of this three-layer structure is continuously varying from the viscosity of inner core $\muin$ to the viscosity of ambient fluid $\muout$. A simple example is the linearly varying $\mumid$, namely, $\mumid(r) = \muin +\frac{\muout - \muin}{\Rout - \Rin}(r - \Rin)$, where $ \Rin < r < \Rout$. The $N$-layer structure (left inset of  \figref{FigContNLayer}) with radius $\myupper{R}{1} = \Rin$, $\myupper{R}{N-1} = \Rout$,  $\Rn = \myupper{R}{1} + \frac{\myupper{R}{N-1} - \myupper{R}{1}}{N-2}(n-1)$,  and viscosity $\myupper{\mu}{1} = \muin$, $\myupper{\mu}{N} = \muout$, and $\mun = \myupper{\mu}{1} + \frac{\myupper{\mu}{N} - \myupper{\mu}{1}}{N-1}(n-1)$ approaches this continuous model for large $N$. In order to obtain a physically realistic continuous-viscosity model with an energy that is both finite and extensive (proportional to volume), we postulate a volume energy density $\gammacont$ analogous to surface energy. We approximate this by an $N$-layer model constructed to have the same total interfacial energy:
\begin{equation}
  \label{eq:eqContEnergy}
  \gammacont  \int_{\Rin}^{\Rout} 2 \pi r \opd r = \sum_{n=1}^{N-1} \gamman  2\pi \Rn,
\end{equation}
 where $\gammacont$ is an appropriate energy (per unit volume) of the inhomogeneity. Corresponding to a uniform $\gammacont$, the surface-tension coefficient in this $N$-layer structure is same on all the interfaces: namely $\gamman = \gamma(N)$ for all $n$. From \eqref{eqContEnergy}, we obtain the equivalent surface tension $\gamma(N)$ in an $N$-layer structure
 \begin{equation}
   \label{eq:eqContGamma}
   \gamma(N) = \frac{\gammacont  \int_{\Rin}^{\Rout} r \opd r}{ \sum_{n=1}^{N-1}\left[\Rin + \frac{\Rout - \Rin}{N-2} (n-1) \right] } = O\left(\frac{1}{N}\right).
 \end{equation}
With the parameters described above, we compute the maximum growth rate $\sigmamax$ and the maximum effective growth rate $\sigmaeffmax$ for this $N$-layer structure. As shown in  \figref{FigContNLayer}, both $\sigmamax$ and $\sigmaeffmax$ approach constants as $N \to \infty$, which should be the corresponding growth rates of the continuous three-layer model. In this example, $\sigmaeffmax$ and $(\sigmamax)^{\rm eff}$ are the same for all $N$. The corresponding eigen-amplitudes $\bolddeltaRmax$ and $\bolddeltaRmaxeff$ (for $N$=70) are plotted in the right inset of  \figref{FigContNLayer}.

\section{Conclusions}
In this paper, we presented a complete linear stability analysis of
concentric cylindrical shells in the Stokes regime (with the Navier--Stokes regime in Appendix~\ref{SecFullNS}) and considered a
few interesting examples and limiting cases.  Many possibilities
present themselves for future work.  First, even in the cylindrical
Stokes regime, only a few combinations of thicknesses and material
properties have been considered so far---it seems quite possible that
consideration of larger parameter spaces, perhaps aided by
computational optimization, could identify additional regimes for
breakup processes, such as competitions between additional
length scales or ``effective'' properties in many-layer systems that
differ substantially from the constituent materials.  Second, one
could extend this work beyond the incompressible Navier--Stokes regime to include fluid compressibility or even other physical phenomena such as
viscoelasticity that may play a role in experiments (for example, fibres
are drawn under tension).  Third, one could consider non-cylindrical
geometries.  This seems especially important in light of the recent
experimental observations of azimuthal breakup in cylindrical
thin-shell fibre structures~\mycite{Deng2008}, since this paper points
out that azimuthal breakup cannot arise in purely cylindrical structures (at least, not from surface tension alone). Instead, one may need to consider the
``neck-down'' structure of the fibre-drawing process, in which a large
preform is pulled to a long strand with a much smaller diameter.  More
generally, such intriguing experimental results indicate that a rich
variety of new instability phenomena may arise in emerging multi-fluid
systems, with corresponding new opportunities for theoretical
analysis.

\begin{acknowledgments}
 We are grateful to Y. Fink and J. Bush at MIT for helpful discussions. X. L thanks Y. Zhang and P. Buchak for kind help on numerical simulations. D. S. D. acknowledges the support and encouragement from M. Bazant. This work was supported by the Center for Materials Science and Engineering at MIT through the MRSEC Program of the National Science Foundation under award DMR-0819762, and by the U.~S. Army through the Institute for Soldier Nanotechnologies under contract W911NF-07-D-0004 with the U.~S. Army Research Office.
\end{acknowledgments}




\appendix

\section{Computations of the curvature}
\label{AppendixCurvature}
Here we derive the curvature terms in \eqref{eqnormal}. The level-set function $\myupper{\phi}{n}(r,z,t) = 0 $ corresponds to the $n$-th interface. The unit outward normal vector of this interface  is 
\begin{equation}
  \label{eq:eqVectornj}
  \nn = (\myupper{n_r}{n}, \myupper{n_z}{n}) = \frac{ \vecnabla \phin}{ | \vecnabla \phin|} = \frac{ (\frac {\partial \phin}{ \partial r}, \frac{\partial \phin}{\partial z} )} { \sqrt { (\frac {\partial \phin}{ \partial r})^2 + ( \frac{\partial \phin}{\partial z} )^2 }} = \frac{ ( 1, -\opi k \deltaRn  \ope^{\opi(kz - \omega t)} )} { \sqrt { 1 + O[(\deltaRn)^2] }}
\end{equation}
and the unit tangential vector is
\begin{equation}
  \label{eq:eqVectortj}
  \tn = (\myupper{n_z}{n}, -\myupper{n_r}{n}) =  \frac{ ( -\opi k\deltaRn  \ope^{\opi(kz - \omega t)} , -1)} { \sqrt { 1 + O[(\deltaRn)^2 ] }}.
\end{equation}
The curvature $\kappan$ can now be computed as
\begin{equation} \label{eq:eqCurvature}
  \begin{split}
  \kappan & = {\vecnabla} \cdot \nn  =  \frac{ \partial \myupper{n_r}{n}}{ \partial r} + \frac{ \myupper{n_r}{n}}{r} + \frac{\partial \myupper{n_z}{n}}{ \partial z} \\
     & =  \frac{1}{ \sqrt{1 + O\left[(\deltaRn)^2\right]}} \frac{1}{  \Rn  + \deltaRn  \ope^{\opi(kz - \omega t)} } + 
    \frac{\deltaRn k^2   \ope^{\opi(kz - \omega t)} }{ \sqrt{1 + O\left[(\deltaRn)^2 \right]} } + O\left[(\deltaRn)^2 \right] \\
 & = \frac{1}{ \Rn } + \deltaRn  \left(k ^2 - \frac{1}{ (\Rn)^2} \right)  \ope^{\opi(kz - \omega t)} +  O\left[(\deltaRn)^2 \right].
  \end{split} 
\end{equation}
The normal velocity of the fluids on the interface must be equal to the normal velocity of that interface, and thus  $\frac{\partial \zetan}{\partial t} = \un \cdot \vec{n}^n$ on the interface $ r = \zetan(z,t)$, where $\un$ is the velocity vector. For the at-rest steady state \eqref{eqb4N} and \eqref{eqb4}, to the lowest order in $\delta R$, this gives
\begin{equation}\label{eq:eqkinematic}
     -\opi \omega \deltaRn = \deltaurn( \Rn ).
 \end{equation}
Note that \eqref{eqkinematic} establishes the relation between the displacement amplitude $\deltaRn$ and the interface velocity $\deltaurn(\Rn)$. Substituting \eqref{eqkinematic} into \eqref{eqCurvature}, we obtain the lowest-order curvature $\kappan$ in terms of the interface velocity $\deltaurn(\Rn)$:
\begin{equation} \label{eq:eqCurvature2}
  \kappan  = \frac{1}{ \Rn } + \frac{ \deltaurn( \Rn )}{-\opi \omega}  \left(k ^2 - \frac{1}{ (\Rn)^2} \right)  \ope^{\opi(kz - \omega t)}. 
\end{equation}

\section{Full 3-dimensional Stokes-flow numerical simulation scheme for coupled cylindrical-shell system}
\label{SecNumScheme}
In this section, we briefly present the numerical scheme that we used in \mysection\ref{SecCompetingModes} to simulate the instabilities of coupled cylindrical-shell systems. We adopt a 3-dimensional Cartesian level-set approach. We use a separate level-set function $\phin$ to denote each interface, and generalize the formulation of \citeasnoun{Chang1996} to $N$ fluids by using $N-1$ level-set functions governed by the following equations:
\begin{equation}
  \label{eq:eqNumStokes}
  - \vecnabla p + \vecnabla \cdot \left[\mu ({\vecnabla} {\vec{U}} + {\vecnabla} {\vec{U}}^{\rm T} ) \right] = \sum_{n=1}^{N-1} \gamman \delta(\phin) \kappa(\phin) \frac{\vecnabla \phin}{|\vecnabla \phin|},
\end{equation}
and
\begin{equation}
  \label{eq:eqNumConvection}
  \frac{\partial \phin}{\partial t} + {\vec{U}} \cdot \vecnabla \phin = 0,
\end{equation}
where $\vec{U}$ is velocity, $p$ is pressure, $\phin = 0$ denotes the interface between the $n$-th and $(n+1)$-th layers, $\gamman$ is the surface-tension coefficient of the $n$-th interface, $\delta(\cdot)$ is a Dirac delta function, $\kappa$ is curvature, and $\mu$ is viscosity. 
 
 The viscosity $\mu$, now defined in the whole coupled system, is
\begin{equation}
  \label{eq:eqNumMu}
  \mu(\vecx ) = \myupper{\mu}{1} + \sum_{n=1}^{N-1} (\munpa - \mun) H(\phin (\vecx)),
\end{equation}
where $H(\cdot)$ is the Heaviside step function. The curvature $\kappa(\phin)$ can be computed directly by $\kappa(\phin) = \vecnabla \cdot \frac{\vecnabla \phin}{ |\vecnabla \phin|} $, since $\frac{\vecnabla \phin}{ |\vecnabla \phin|}$ is the unit outward normal vector of the $n$-th interface. 

Given the level-set function $\phin(\vecx,t)$ at time $t$, we first solve the steady Stokes equations \eqref{eqNumStokes} to obtain the velocity ${\vec{U}}(\vecx,t)$. With the known velocity ${\vec{U}}(\vecx,t)$ at time $t$, the level-set function $\phin(\vecx,t + \Delta t)$ can be obtained by solving the convection equation \eqref{eqNumConvection}. 

In our implementation, the computation cell is a box with dimensions $a \times a \times \ell$ in Cartesian coordinates, with periodic boundary conditions. We choose $a$ and $\ell$ large enough such that the periodicity does not substantially affect the breakup process. We solve equations~\eqref{eqNumStokes} by a spectral method: we represent $\vec{U}$ and $p$ by Fourier series (discrete Fourier transforms). For the constant $\mu$ case of \mysection\ref{SecCompetingModes}, \eqref{eqNumStokes} is diagonal in Fourier space and can be solved in one step by fast Fourier transforms (FFTs). More generally, for variable viscosity, we find that an iterative solver such as GMRES (generalized minimal residual method) or BiCGSTAB (biconjugate gradient stabilized method) \mycite{Barrett1994}  converges in a few iterations with a constant $\mu$ preconditioner (i.e., block Jacobi) using the average $\mu$. The level-set functions are described on the same grid, but using finite differences: WENO (weighted essentially non-oscillatory: \citeasnoun{Liu1994}) in space, third order TVD (total variation diminishing) Runge-Kutta method in time \mycite{Shu1989}. The $\delta(\cdot)$ function is smoothed over 3 pixels with a raised-cosine shape~\mycite{Osher2002}. We use the reinitialization scheme of \citeasnoun{Sussman1994} to preserve the signed distance-function property $|\vecnabla \phin| = 1$ of $\phin$ after each time step.

\begin{figure}
\centering
 \includegraphics[width=0.7\textwidth]{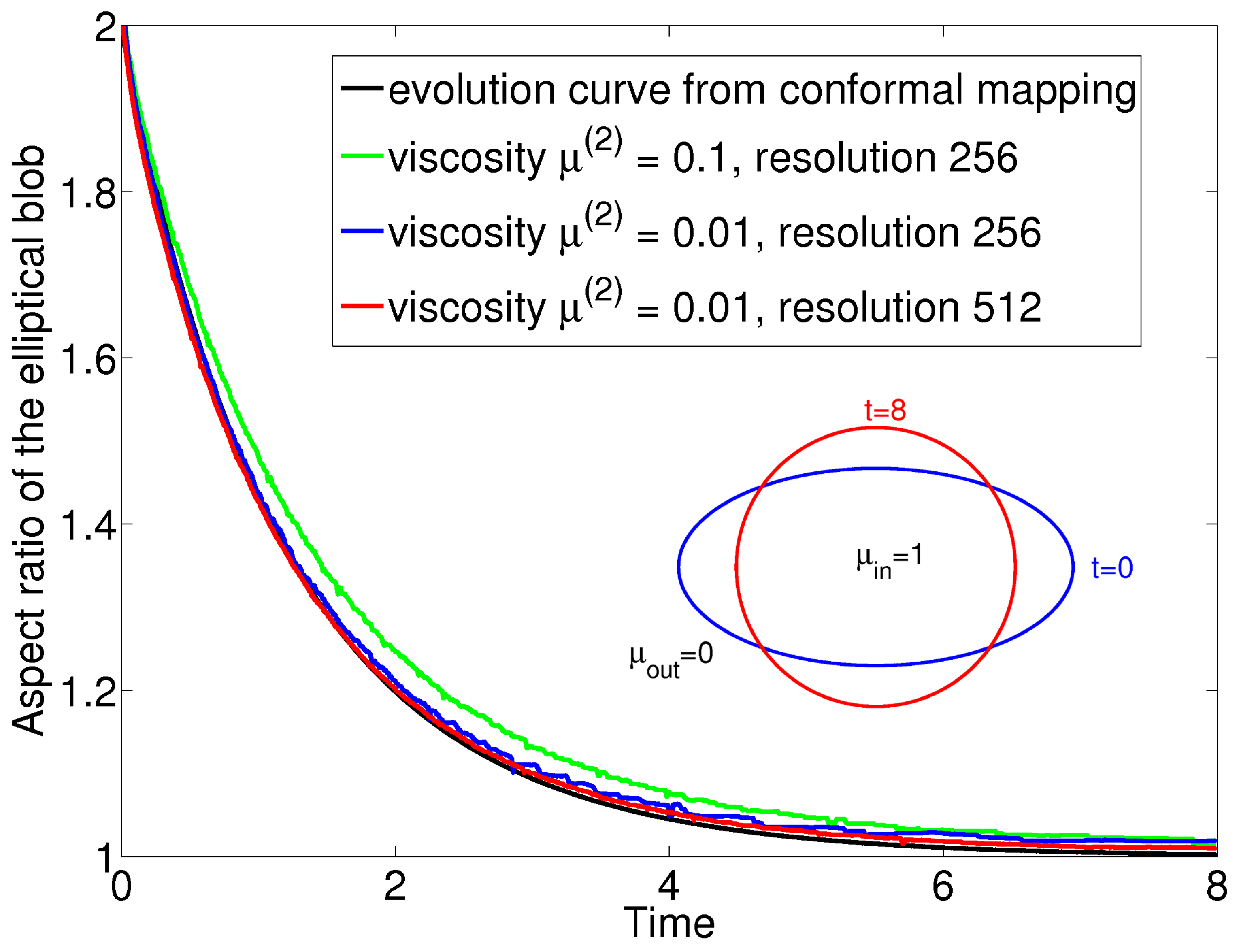} 
\caption{The aspect ratio of a 2-dimensional elliptical blob versus time, obtained by different methods and implementations. For the system initially bounded by $x^2/4 + y^2= 1$, with the elliptical blob viscosity $\muin=1$,  the ambient fluid viscosity $\muout=0$ and the surface tension $\gamma=1$, \citenameasnoun{Buchak2010} used the conformal mapping method via finite element implementation, obtaining the black evolution curve. The evolution curves (green, blue and red) given by our simulations converge to the black curve  with the increasing resolutions and as the ambient viscosity $\myupper{\mu}{2}$ goes to zero.
\label{fig:FigEllipse}}
\end{figure}

Our simulation code is validated against a well-studied case: the evolution of a 2-dimensional elliptical blob \mycite{Kuiken1990, Hopper1991, Tanveer1995, Crowdy2002, Crowdy2003}. It is known that the plane Stokes flow, initially bounded by a simple smooth closed elliptic curve, will eventually become circular under the effect of surface tension. \citenameasnoun{Crowdy2002} illustrated that the evolution via a series of ellipse shapes is remarkably good approximation to the dynamics of a sintering ellipse [even though \citenameasnoun{Hopper1991} showed that the exact evolution shapes are not strictly elliptical]. Suppose the plane Stokes flow is bounded by the ellipse $x^2/4 + y^2 =1$ at $t=0$. The viscosity of the elliptical blob $\muin=1$,  the viscosity of ambient fluid $\muout=0$, and the surface tension $\gamma=1$. \citenameasnoun{Buchak2010} implemented the conformal mapping method \mycite{ Tanveer1995, Crowdy2002, Crowdy2003} and computed the evolution of the boundaries. The aspect ratio (the major axis over the minor axis) of the ellipses versus time is plotted (black curve) in \figref{FigEllipse}. Since our simulation code only works for non-zero $\mun$, the evolution under $\myupper{\mu}{1}=1$ and $\myupper{\mu}{2} \to 0$ is expected to converge to the black evolution curve obtained by the conformal mapping. In \figref{FigEllipse}, we also plotted the evolution curves from our method with $\myupper{\mu}{2}=0.1$ and resolution $256\times 256$ (green curve), $\myupper{\mu}{2}=0.01$ and resolution $256\times 256$ (blue curve), and $\myupper{\mu}{2}=0.01$ and resolution $512\times 512$ (red curve). With high resolutions and small $\myupper{\mu}{2}$, the evolution curves obtained by our simulation codes converge to the one given by a different method with an independent implementation.

\section{Linear stability analysis for concentric fluid shells governed by the full Navier--Stokes equations}\label{SecFullNS}
In this section, we extended our linear stability analysis to concentric cylindrical fluid shells governed by the full Navier--Stokes equations. Let $\rhon$ and $\mun$ denote the density and viscosity of the $n$-th fluid; $\urn (z)$ is the radial component of the velocity and $\uzn (z)$ is the axial component of the velocity. Following a linear stability analysis similar to ~\mysection\ref{sec:PerturbedState}, we find that the pressure $\pn(z)$ in the $n$-th fluid still satisfies Laplace's equation~\eqref{eqLaplace}. Therefore, the perturbed pressure still satisfies the modified Bessel equation~\eqref{eqb10} and the solution in~\eqref{eqb11} is still valid. The velocity is obtained by solving the linearized Navier--Stokes equations
\begin{equation}\label{eq:eqLinearFullNSa}
-\rhon \frac{\partial \urn}{\partial t} +  \mun \left( \frac{\partial^2 \urn}{  \partial r^2} + \frac{1}{r} \frac{\partial \urn}{\partial r} - \frac{\urn}{r^2} + \frac{\partial^2 \urn}{\partial z^2} \right) =  \frac{ \partial \pn}{ \partial r} 
\end{equation}
and
\begin{equation}\label{eq:eqLinearFullNSb}
- \rhon \frac{\partial \uzn}{\partial t} + \mun \left( \frac{\partial^2 \uzn}{  \partial r^2} + \frac{1}{r} \frac{\partial \uzn}{\partial r} + \frac{\partial^2 \uzn}{\partial z^2} \right) =   \frac{ \partial \pn}{ \partial z} .
\end{equation}
Note that the nonlinear convection terms do not appear in the linearized equations \eqref{eqLinearFullNSa} and \eqref{eqLinearFullNSb} because the basic steady state \eqref{eqb4N}--\eqref{eqb4} is at rest. Substituting the perturbed pressure \eqref{eqb11} into equations \eqref{eqLinearFullNSa}--\eqref{eqLinearFullNSb}, we find that the radial component of the perturbed velocity in \eqref{eqb12} is replaced by
\begin{multline}
  \deltaurn (r) = \cna \frac{K_1(kr) - K_1(\kn r) }{-\opi \omega \rhon /k} - \cnb  \frac{I_1(kr) - I_1(\kn r) }{-\opi \omega \rhon /k} \\
+ \cnc  \frac{ K_1(\kn r)}{2 \mun k} + \cnd  \frac{I_1(\kn r)}{2 \mun k}
\end{multline}
and the axial component of the perturbed velocity in \eqref{eqb13} now becomes
\begin{multline}
 \deltauzn (r) = \cna \frac{ K_0(kr) - \lambdan K_0 (\kn r)}{\rhon \omega / k} + \cnb \frac{I_0(kr) - \lambdan I_0(\kn r)}{\rhon \omega /k} \\ - \cnc \frac{\opi \lambdan K_0(\kn r)}{2 \mun k} + \cnd \frac{\opi \lambdan I_0(\kn r) }{2 \mun k},
\end{multline}
where
\begin{equation}
\label{eq:eqLambdan}
\lambdan = \sqrt{1+  \frac{-\opi \omega\rhon}{ \mun k^2}} \qquad \mbox{and} \qquad \kn = \lambdan k. 
\end{equation}
After matching boundary conditions \eqref{eqContRadial}, \eqref{eqContAxial}, \eqref{eqtangential2} and \eqref{eqnormal2}, we can obtain the dispersion relation by solving the same determinant equation \eqref{eqDeterminant}, except that $\matAnnp$ and $\matBn$ from \eqref{eqMatrixAn} and \eqref{eqMatrixBn} are replaced by
 {\myfontsize
 \begin{multline} \label{eq:eqFullNSMatrixAn}
   \matAnnp_{\mbox{ns}}  =  \\
  \begin{bmatrix}
 \frac{\Kan - \Ka[\knpr \Rn] }{\rhonpr / 2 k^2} & -\frac { \Ian - \Ia[\knpr \Rn]}{\rhonpr / 2 k^2} &  \frac{\Ka[\knpr \Rn]}{\munpr} &  \frac{\Ia[\knpr \Rn]}{\munpr} \\
 \frac{\Kzn - \lambdanpr \Kz[\knpr \Rn]}{-\rhonpr / 2 k^2 } & \frac {\Izn - \lambdanpr \Iz[\knpr \Rn] }{-\rhonpr / 2 k^2} & \frac{\lambdanpr \Kz[\knpr \Rn]}{-\munpr} &  \frac{ \lambdanpr \Iz[\knpr \Rn]}{\munpr }\\
\frac{\Kan - \alphanpr \Ka[\knpr \Rn]}{ \rhonpr / 2 \munpr k^2} & \frac{\Ian - \alphanpr \Ia[\knpr \Rn] }{-\rhonpr / 2 \munpr k^2} & \alphanpr \Ka[\knpr \Rn] & \alphanpr \Ia[\knpr \Rn]\\
\triangle_1 & \triangle_2 & \triangle_3 & \triangle_4
 \end{bmatrix},
\end{multline}}
where
\begin{equation}
  \label{eq:eqAlphanpr}
  \alphanpr = 1 + \frac{-\opi \omega \rhonpr}{2 \munpr k^2}
\end{equation}
\begin{equation}
  \triangle_1 = \frac{ \alphanpr \Kzn + \Kan/k \Rn}{\rhonpr / 2\munpr k^2} - \frac{\lambdanpr \Kz[\knpr \Rn] + \Ka[\knpr \Rn]/k \Rn }{\rhonpr / 2 \munpr k^2}
\end{equation}
\begin{equation}
  \triangle_2 = \frac{ \alphanpr \Izn - \Ian /k \Rn }{ \rhonpr / 2 \munpr k^2} + \frac{ -\lambdanpr \Iz[\knpr \Rn] + \Ia[\knpr \Rn]/k \Rn  }{ \rhonpr / 2 \munpr k^2}
\end{equation}
\begin{equation}
  \triangle_3 =  \lambdanpr  \Kz[\knpr \Rn] + \Ka[\knpr \Rn]/k \Rn
\end{equation}
\begin{equation}
  \triangle_4 =  -\lambdanpr \Iz[\knpr \Rn] + \Ia[\knpr \Rn]/k \Rn
\end{equation}
and
\begin{multline}\label{eq:eqFullNSMatrixBn}
   \matBn_{\mbox{ns}}  =  \frac{ - \gamman k \left[ 1- \frac{1}{(k \Rn)^2} \right]}{ 2  } \\ \times
  \begin{bmatrix}
   0 &0 &  0 &  0 \\
   0 &0 &  0 &  0 \\
   0 &0 &  0 &  0 \\
   \frac{\Kan -\Ka[\kn \Rn] }{\rhon /2 k^2} & -\frac{\Ian -\Ia[\kn \Rn] }{\rhon / 2k^2} & \frac{ \Ka[\kn \Rn]}{\mun}  &  \frac{\Ia[\kn \Rn]}{\mun}
  \end{bmatrix}.
\end{multline}

Note that, because of the $\omega$ in $\lambdan$ and $\kn$, the matrix $\MN$ becomes nonlinear in $\omega$ (or $1/\omega$), and can no longer be reduced to a generalized eigenproblem. Instead, one must solve the nonlinear eigenproblem $\myupper{\matrx{M}_2}{N}(k, \omega) \boldsymbol {\xi} = \opi \omega \myupper{\matrx{M}_1}{N}(k, \omega)\boldsymbol {\xi} $. Numerous methods have been developed for such problems~\mycite{Andrew1995, Guillaume1999, Ruhe2006, Voss2007, Liao2010}.

The formulae \eqref{eqMatrixAn} and \eqref{eqMatrixBn} of $\matAnnp$ and $\matBn$ for Stokes flow can be obtained directly from the formulae \eqref{eqFullNSMatrixAn} and \eqref{eqFullNSMatrixBn} of  $\matAnnp_{\mbox{ns}}$ and $\matBn_{\mbox{ns}}$ for general flow by taking the limit $\rhon \to 0$. As mentioned in~\mysection~\ref{SecTomotikaCase}, the most straightforward formulation of the Navier--Stokes matrices yields dependent columns when $\rho \to 0$.  Here, we have chosen an appropriate linear combination of columns to avoid this difficulty, which is equivalent to the procedure suggested by \citeasnoun{Tomotika1935}.

However, the corresponding formulae for inviscid fluids cannot be obtained by simply taking the limit $\mun \to 0$. For any small but nonzero $\mun$, the current formulation takes into account the boundary-layer effects \mycite{Batchelor1973} by imposing a no-slip condition. For inviscid flows, we cannot assume that the axial velocities are continuous across the interfaces, since no-slip boundary conditions are not applied. Whenever the no-slip boundary condition is not applied, one has an additional degree of freedom, the equilibrium-state velocities $\baruzn$ of the layers. This is easily incorporated, because it merely converts several $\omega$ expressions to $\omega - \baruzn k$.  This happens in two places.  First, the velocity adds an additional inertial term $ -\rhon \baruzn {\partial \urn}/{\partial z}$ to the left side of~\eqref{eqLinearFullNSa} and $ -\rhon \baruzn \frac{\partial \uzn}{\partial z}$ to the left side of~\eqref{eqLinearFullNSb}.  Second, it adds a new first-order term to equation~\eqref{eqFullNSnormal} for continuity of normal velocity, since there is a term from $\baruzn$ multiplied by the $\deltaRn$ in the numerator of \eqref{eqVectornj} for the normal vector.  These terms change $\omega$ to $\omega-\baruzn k$ in \eqref{eqLambdan} for $\lambdan$ and to $\omega-\baruznp k$ in \eqref{eqAlphanpr} for $\alphanpr$, and they also multiply every $\matBn_{\mbox{ns}}$ matrix (including $\myupper{\tilde{\matrx{B}}}{1}_{\mbox{ns}}$) by $\omega/(\omega-\baruzn k)$ [cancelling the $1/\omega$ factor multiplying $\MbN$ in \eqref{eqMatrixM}]; the $\matAnnp_{\mbox{ns}}$ and $\matBn_{\mbox{ns}}$ matrices are otherwise unchanged, since $\baruzn$ must be equal for adjacent viscous layers.  If the $n$-th fluid is inviscid while the $(n - 1)$-th and/or $(n + 1)$-th fluids are viscous, then $\myupper{\matrx{A}}{n-1,n-1}_{\mbox{ns}}$ and $\myupper{\matrx{B}}{n-1}_{\mbox{ns}}$, and/or $\myupper{\matrx{A}}{n,n+1}_{\mbox{ns}}$, respectively, become $3 \times 4$ matrices that can be obtained from \eqref{eqFullNSMatrixAn} and \eqref{eqFullNSMatrixBn} by eliminating the second row (corresponding to continuity of the tangential component of the velocity).  If the $n$-th layer is inviscid, regardless of the adjacent layers, $\matAnpn_{\mbox{invsd}}$ and $\matBn_{\mbox{invsd}}$ are $3 \times 2$ matrices: not only has continuity of the tangential component of the velocity disappeared, but also the $\partial/\partial r$ derivatives of the velocities in the momentum equations \eqref{eqLinearFullNSa} and \eqref{eqLinearFullNSb} disappear when $\mun=0$, eliminating the $\myupper{c_{\{3,4\}}}{n}$ degrees of freedom. (This eliminates the need for the linear combinations of columns mentioned above, further simplifying these matrices.) More explicitly, the $\matAnpn_{\mbox{invsd}}$ and $\matBn_{\mbox{invsd}}$ matrices for an inviscid $n$-th layer are obtained by matching the boundary conditions \eqref{eqFullNSnormal}, \eqref{eqtangential2}, and \eqref{eqnormal2}, giving
\begin{equation}
    \matAnpn_{\mbox{invsd}}  = 
  \begin{bmatrix}
 \frac{\omega - \baruznp k}{\omega - \baruzn k}\frac{\Kanpr }{\rhon / 2 k^2} & -\frac{\omega - \baruznp k}{\omega - \baruzn k} \frac{ \Ianpr}{\rhon / 2 k^2}  \\
 0 & 0 \\
-\opi (\omega - \baruzn k) \Kznpr & -\opi (\omega - \baruzn k)  \Iznpr
 \end{bmatrix} 
\end{equation}
\begin{equation}
  \matBn_{\mbox{invsd}}  =\frac{\omega}{\omega - \baruzn k} \frac{ - \gamman k \left[ 1- \frac{1}{(k \Rn)^2} \right]}{ 2 }  
  \begin{bmatrix}
 0 & 0 \\
 0 & 0 \\
 \frac{\Kan }{\rhon /2 k^2} & -\frac{\Ian}{\rhon / 2k^2}
 \end{bmatrix}.
\end{equation}
For the special case $\baruzn = 0$ (at-rest steady state), the above formulae are equivalent to the ones obtained by eliminating the second row, eliminating the third and fourth columns, then taking the limit $\mun \to 0$ in the general formulae \eqref{eqFullNSMatrixAn} and \eqref{eqFullNSMatrixBn}. More precisely, for the viscosity term in \eqref{eqLinearFullNSa} and \eqref{eqLinearFullNSb} to be negligible, one must have $\mun \ll~\omega \rhon / k^2$. (Note that this is length- and time-scale dependent, so the validity of neglecting viscosity terms depends on the $\omega$ and $k$ of the dominant growth mode.)  It is also interesting to consider a Galilean transformation in which a constant $\barvz$ is added to $\baruzn$ for all $n$, which cannot change the physical results. Here, because all $\omega$ factors are accompanied by $-\baruzn k$, it is clear that such a transformation merely shifts all of the mode frequencies $\omega_j(k)$ by $\barvz k$, which does not change the growth rates (the imaginary part), while the shift in the real frequency is simply due to the frequency-$k$ spatial oscillations moving past any fixed $z$ at velocity $\barvz$.

As a validation check, we find that our formulation gives the same dispersion relations for various Navier--Stokes cases discussed in the previous literature: e.g., a single inviscid jet in air (ignoring the air density and viscosity) \mycite{Rayleigh1879}, a single viscous jet in air (ignoring the air density and viscosity) \mycite{Rayleigh1892}, a single viscous jet with high velocity in air (considering the air density but ignoring the viscosity) \mycite{Sterling1975} and a compound jet in air (ignoring the air density and viscosity) \mycite{Chauhan2000}.

\bibliographystyle{jfm2}
\bibliography{InstabilityRef}

\end{document}